\documentclass{jfm}

\usepackage{graphicx}
\usepackage{natbib}
\usepackage{epstopdf}
 \usepackage{amsmath,amssymb}
\usepackage{floatrow}
\usepackage{subfig}
\usepackage{multirow}
\usepackage{slashbox}

\def\CO2{CO$_2$}
\newcommand{\param}{\Pi}
\newcommand{\perm}{K}
\newcommand{\Td}{\mathcal{T}_d}
\newcommand{\Ta}{\mathcal{T}_{a}}
\newcommand{\Tad}{\mathcal{T}_{ad}}

\newcommand{\U}{\mathcal{U}}
\renewcommand{\H}{\mathcal{H}}

\newcommand{\coeff}{\alpha}
\newcommand{\coeffup}{\alpha_{\tiny\mathrm{u}}}
\newcommand{\expo}{\beta}

\newcommand{\lpls}{\hat}

\shorttitle{Convective \CO2 dissolution in a closed porous media system}
\shortauthor{B. Wen, D. Ahkbari, L. Zhang and M. Hesse}

\title{Dynamics of convective carbon dioxide dissolution in a closed porous media system}

\author{Baole Wen\aff{1}, Daria Ahkbari\aff{2}, Li Zhang\aff{3} \and  Marc A. Hesse\aff{1,2} \corresp{\email{mhesse@jsg.utexas.edu}}}

\affiliation{\aff{1} Institute for Computational Engineering and Sciences, University of Texas at Austin,
Austin, TX 78712, USA, \aff{2} Department of Geological Sciences, University of Texas at Austin, Austin, TX, 78712, US, \aff{3} Department of Engineering Mechanics and CNMM, Tsinghua University, Beijing, 100084, China, }

\pubyear{2010}
\volume{650}
\pagerange{119--126}
\date{?; revised ?; accepted ?. - To be entered by editorial office}
\begin{document}

\maketitle

\begin{abstract}
Motivated by geological carbon dioxide (\CO2) storage, many recent studies have investigated the fluid dynamics of solutal convection in porous media. Here we study the convective dissolution of \CO2 in a closed system, where the pressure in the gas declines as convection proceeds. This introduces a negative feedback that reduces the convective dissolution rate even before the brine becomes saturated. We analyse the case of an ideal gas with a solubility given by Henry's law, {\color{black}{in the limits of very low and very high Rayleigh numbers}}. The equilibrium state in this system is determined by the dimensionless dissolution capacity, $\param$, which gives the fraction of the gas that can be dissolved into the underlying brine.  Analytic approximations of the pure diffusion problem with $\param>0$, show that the diffusive base state is no longer self-similar and that diffusive mass transfer declines rapidly with time. Direct numerical simulations {\color{black}{at high Rayleigh numbers}} show that no constant flux regime exists for $\param > 0$; nevertheless, the quantity $F/C_s^2$ remains constant, where $F$ is the dissolution flux and $C_s$ is the dissolved concentration at the top of the domain.  Simple mathematical models are developed to predict the evolution of $C_s$ and $F$ for {\color{black}high-Rayleigh-number} convection in a closed system. The negative feedback that limits convection in closed systems may explain the persistence of natural \CO2 accumulations over millennial timescales.

\end{abstract}


\section{Introduction}\label{sec:introduction}

One promising means of reducing the atmospheric emissions of carbon dioxide (\CO2) is to store it in deep geological formations \citep{Metz2005,Orr2009}.  When \CO2 is injected into a saline aquifer, it forms an immiscible CO$_2$-rich vapour phase which is lighter than the aqueous brine and accumulates at the top of the storage formation. The \CO2 dissolves into the underlying brine and forms a diffusive boundary layer beneath the gas water contact.
The brine density increases with aqueous \CO2 concentration and the boundary layer can become unstable and lead to convective overturn within the brine \citep{Weir1995,Ennis-King2005}. Convective mass transfer can greatly increase the dissolution rate of the injected buoyant \CO2 vapour and hence contributes to safe long-term storage \citep{Neufeld2010,Sathaye2014}. 

\begin{figure}[t!]
{
  \centering
  \includegraphics[width=0.8\textwidth]{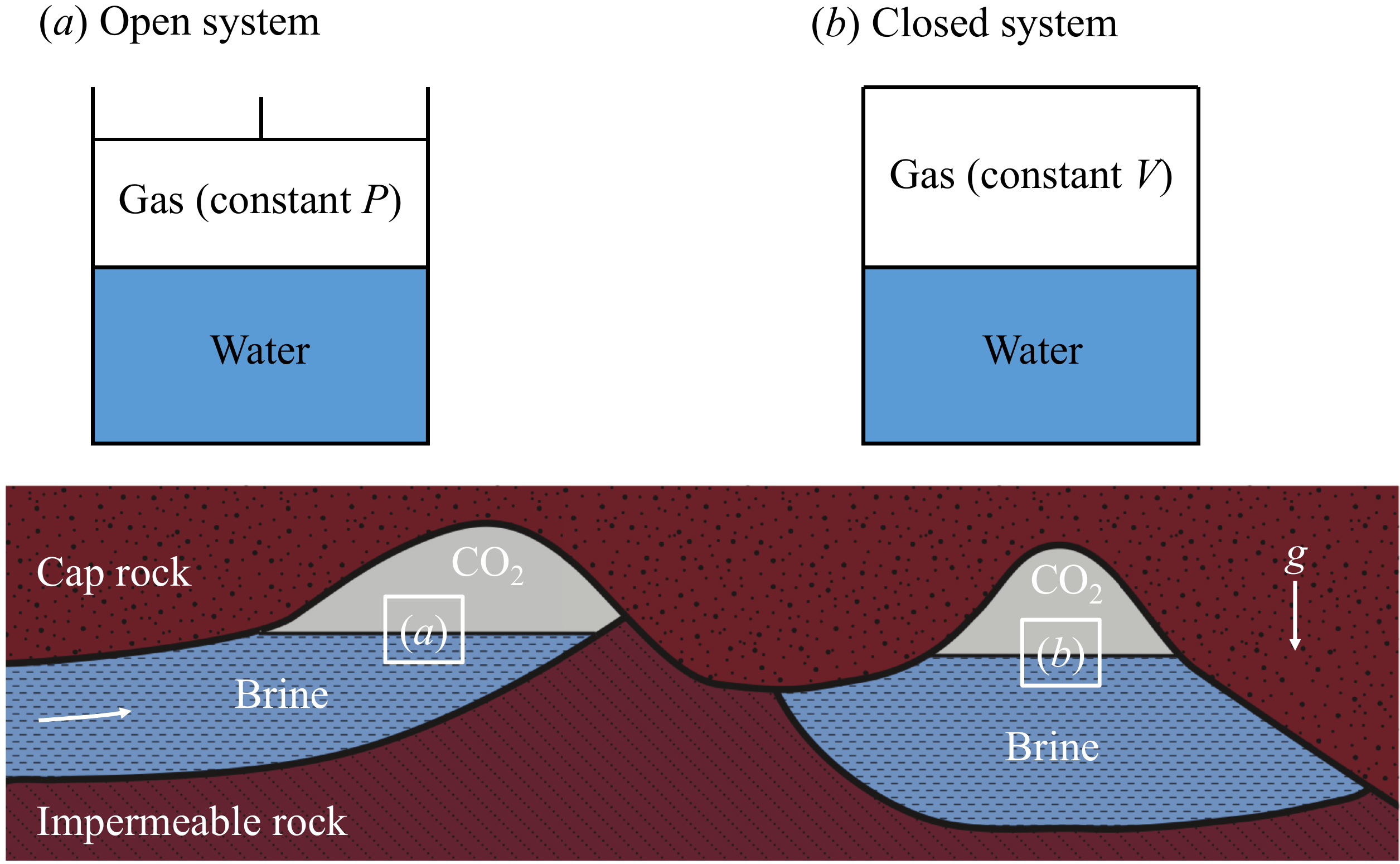}
}
\caption{Schematic showing natural CO$_2$ reservoirs with and without structural closure. In system ($a$), the brine can easily move laterally due to the open structure: as CO$_2$ dissolves in the brine, new brine will be filled into the reservoir, thereby rising the CO$_2$-brine interface and keeping the gas pressure constant; in system ($b$), no brine can escape from or be filled into the reservoir  due to the structural closure so that the interface is fixed and the dissolution of CO$_2$ will reduce the gas pressure.  We refer to the systems ($a$) and ($b$) as open and closed systems, respectively, and the corresponding idealised rectangular systems are shown at the top.}  \label{fig:Schematic_open_closed}
\end{figure}

This application has motivated a large amount of recent work on convection in porous media \citep{Huppert2014, RiazCinar2014, Emami-Meybodi2015b}. Work in fluid dynamics has focused on a simplified model problem that considers convection in the brine driven by a constant concentration applied at the top of the domain. At high Rayleigh numbers, the mass transfer is generally characterised by the succession of three dynamic regimes: an initial diffusive decline until the boundary layer becomes unstable, followed by convective dissolution at constant rate, and finally a rapid decline in dissolution rate as the brine saturates and convection shuts down. Most work has focused on determining the onset of convection \citep{Ennis-King2005,Riaz2006,Hassanzadeh2006, Xu2006}, the convective dissolution rate \citep{Neufeld2010,Pau2010,Hidalgo2012,Hewitt2012}, and the shut down of convection \citep{Slim2010,Hewitt2013shutdown,Slim2013}.

A geological storage site can either be an open or a closed system (see figure~\ref{fig:Schematic_open_closed}). Open sites are typically laterally extensive and allow the compensation of pressure changes by brine migration. In an open system, \CO2 dissolution typically leads to a reduction in the volume of the \CO2 vapour over time, while the \CO2 pressure remains approximately constant due to inflow of brine. The aqueous \CO2 concentration beneath the gas-water contact and therefore the density difference driving convective dissolution remain constant.  In an open system an infinite volume of brine is available, so that all injected \CO2 dissolves eventually. Convective dissolution in an open system therefore proceeds at constant rate until the dense \CO2 saturated fingers begin to interact with the base of the aquifer and dissolution becomes limited by lateral \CO2 transport \citep{Szulczewski2013,Unwin2016}.

Closed sites are typically fault bounded and do not allow compensation of pressure changes by brine migration. Therefore, the volume of \CO2 vapour in a closed system remains essentially constant over time and consequently \CO2 dissolution reduces the pressure in the vapour phase \citep{Akhbari2017}. This leads to a decline of the aqueous \CO2 concentration beneath the gas-water contact and therefore reduces the density difference driving convective dissolution. In addition, the volume of brine in a closed system is finite and may further limit the dissolution into the brine. Convective dissolution of \CO2 in a closed system is therefore limited by both the pressure drop in the vapour and the saturation of the underlying brine. Previous studies of convection in a closed system have focused on the latter \citep{Slim2010,Hewitt2013shutdown,Slim2013}. Here we show that the pressure drop in the gas can limit \CO2 dissolution long before saturation of the brine becomes a limiting factor. These negative feedbacks in closed systems are common in experiments on \CO2 dissolution \citep{Farajzadeh2009,Moghaddam2012,Mojtaba2014,Shi2017} and in some natural \CO2 reservoirs that serve as analogs for geological \CO2 storage \citep{Akhbari2017}.

{\color{black}{Engineered geological storage sites are typically selected such that \CO2 is supercritical to maximise the storage capacity \citep{Orr2009}. However, it is remarkable that many natural \CO2 reservoirs in the continental U.S. are at pressures significantly less than hydrostatic and contain \CO2 in a gaseous state \citep{Akhbari2017}. In particular, this is the case for the Bravo Dome natural \CO2 reservoir which is commonly considered as an analog for engineered \CO2 storage \citep{Broadhead1987, Broadhead1990, Gilfillan2008, Gilfillan2009, Sathaye2014}. Therefore, to simplify the analysis and emphasise the essential new feedback }} we assume that phase behaviour in the closed system is ideal. {\color{black}{However, we have used the same modelling approach to describe high-pressure dissolution experiments with supercritical \CO2 in \citet{Shi2017}, so that the analysis presented here is  not limited to the ideal case.}} Below we give units to avoid confusion that can arise from multiple definitions used for the Henry's law constant. The \CO2 vapour is assumed to be an ideal gas, so that
\begin{eqnarray}
	P_g^*V_g^* = n_gRT, \label{IdealGas}
\end{eqnarray}
where $P_g^*$ [Pa] is the gas pressure, $V_g^*$ [m$^3$] is the gas volume, $n_g$ [mol] is the amount of gas in moles, $R$ [kg m$^2$ /(s$^2$ K mol)] is the universal gas constant, and $T$ [K] is the absolute temperature. The aqueous solution is dilute, so that the local equilibrium between this gas and the dissolved aqueous \CO2 \emph{at the gas-water contact} is given by Henry's law
\begin{eqnarray}
	C_{s}^* =  P_g^*K_h, \label{Henry}
\end{eqnarray}
where $C_{s}^*$ [mol/m$^3$] is the dissolved gas concentration and $K_h$ [mol/(m$^3$ Pa)] is the Henry's law solubility constant. The amount of \CO2 dissolved into a volume of water, $V_w^*$ [m$^3$], in equilibrium with the gas is therefore given by $n_w = V_w^* K_h P_g^*$ [mol].  {\color{black}{Since our study is performed in an closed system, the total volume, i.e. $V_g^* + V_w^*$, and the total amount of \CO2, i.e. $n_g + n_w$, remain constant. We note that our analysis, ignores the slight change in water volume upon \CO2 dissolution as well as the evaporation of water into the gas, both of which are negligible \citep{Shi2017}.}}

Consider a closed system that is initially out of equilibrium and contains a gas with a pressure $P_{g,0}^*$ in contact with a finite volume of water containing no dissolved gas. Once the system reaches \emph{equilibrium}, the normalised final gas pressure and dissolved concentration are given by
\begin{align}
\frac{P_{g,e}^*}{P_{g,0}^*} &= \frac{C_{s,e}^*}{C_{s, 0}^*} = \frac{n_{g,e}}{n_{g,e}+n_{w,e}} = \frac{1}{1+n_{w,e}/n_{g,e}}, \label{eqbm}
\end{align}
where the subscript `$e$' denotes the final equilibrium state, and $C_{s,0}^*=K_hP_{g,0}^*$ is the dissolved concentration at the interface in local equilibrium with the initial pressure. We define the ratio of dissolved to gaseous \CO2 molecules at global equilibrium as a new dimensionless parameter
\begin{align}\label{eq:new_param}
\param = \frac{n_{w,e}}{n_{g,e}}=\frac{V_w^*}{V_g^*}K_hRT.
\end{align}
This \emph{dissolution capacity} is a new dimensionless parameter governing both diffusive and convective mass transport in an ideal closed system.  The pressure drop in a closed system increases with the dissolution capacity. In the limit of small $\param$, the pressure drop in the gas becomes negligible and open system behaviour {\color{black}{(i.e. constant $C_{s}^*$)}} is recovered.  {\color{black}{In following sections, we therefore refer to the system with $\param = 0$ as an open system.}} 




The reminder of this paper is organised as follows.  In the next section, we formulate the dimensional model of convection in the closed porous media system, non-dimensionalize the governing equations, and describe the numerical method to solve these dimensionless equations.  In \S\,\ref{sec:Diffusion_onset}, we give analytic solutions for diffusion in a closed system at early and late times and then investigate the effect of $\param$ on the onset of convection using direct numerical simulations (DNS).  In \S\,\ref{sec:DNS_model}, DNS results are  reported for high-Rayleigh-number solutal convection in closed systems, and the corresponding mathematical models of various dissolution qualities are developed for both the quasi-steady convective  and the shut-down regimes. {\color{black}{In \S\,\ref{sec:discussion}, we use our models to estimate the dissolution process in reservoirs with typical parameter values obtained from geological storage sites, and show some moderate-Rayleigh-number DNS results to more comprehensively understand the dynamics of \CO2 dissolution in Bravo Dome natural gas reservoir. Finally, we summarise the key results in \S\,\ref{sec:Conclusion}.}}

\section{Problem formulation}\label{sec:FORMULATION}

\subsection{Dimensional equations}\label{sec:Dimen_Eqns}
\begin{figure}[t]
    \center{\includegraphics[width=0.7\textwidth]{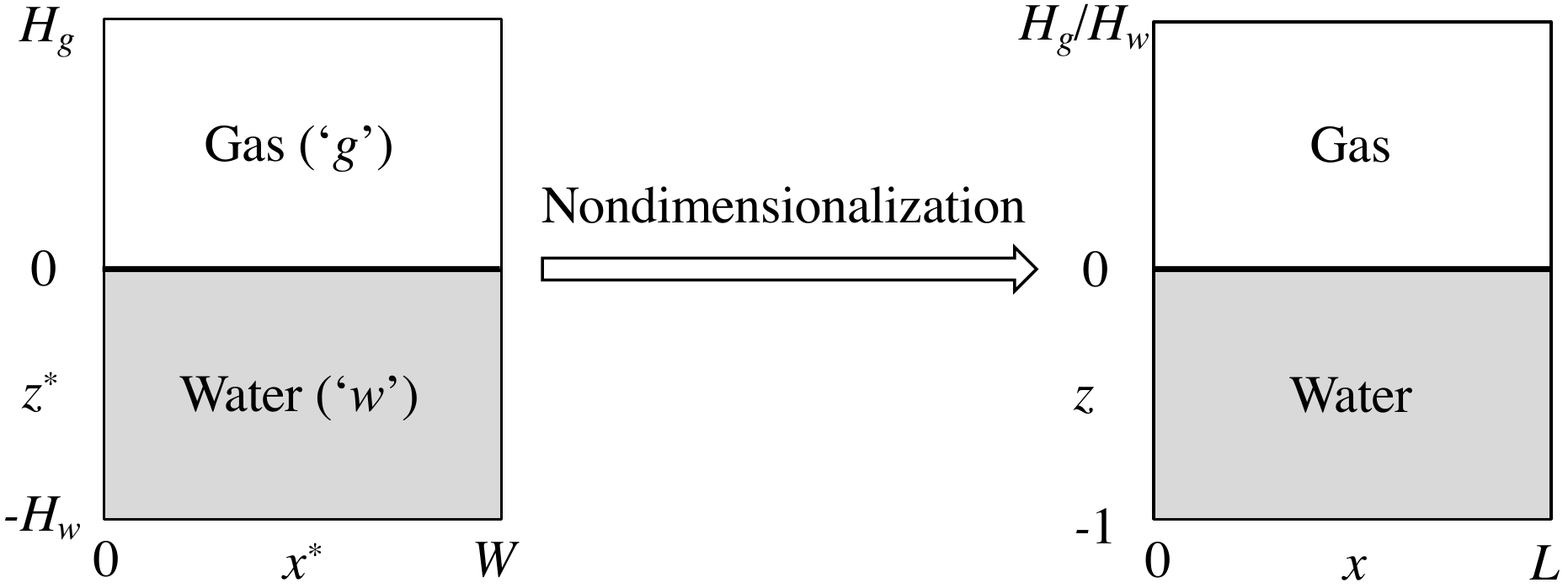}}
    \caption{Geometry for 2D closed porous media system.  The dimensional periodic domain (left) has heights $H_g$ and $H_w$ for the gas and water fields, respectively, and width $W$.  $L = W/H_w$ is the domain aspect ratio for the water field. The continuous dissolution of the gas into the underlying water reduces the gas pressure and then decreases the saturated concentration of water.}  \label{fig:Geometry}
\end{figure}

Consider a two-dimensional (2D), homogeneous, and isotropic porous medium containing gas overlying water (see figure~\ref{fig:Geometry}).  In the limit of negligible capillary forces, the phases are segregated by buoyancy and separated by a sharp interface at $z^*=0$ \citep{Golding2011,Martinez2016}. Therefore, the upper part of the domain, $0 < z^* < H_g$, is occupied only by gas and the lower part, $-H_w < z^* < 0$, is occupied solely by water. Instead of a laterally closed domain we consider a $W$-periodic domain to simplify the DNS in \S\,\ref{sec:Numerical}. In terms of the overall mass balance the periodic system is identical to the closed system.

We assume the gas is ideal and well-mixed, so that the pressure is uniform and given by (\ref{IdealGas}). {\color{black}{The water is incompressible and Boussinesq approximation is valid. We neglect the volume change of water due to the \CO2 dissolution \citep{Shi2017}, so that the domains containing water and gas are fixed, $H_g$ and $H_w$ are constant, and the interface remains at $z^*=0$.}} The system is closed, so that gas and water are coupled through a global mass balance and the local chemical equilibrium along the interface, given by (\ref{Henry}).  Therefore, the governing equations for convection in a closed system comprise  a system of partial differential equations (PDE's) describing the convective mass transport in the water and an ordinary differential equation (ODE) for the evolution of the gas. The ODE is coupled to the system of PDE's though the mass flux, $F^*$, across the interface.

The solute-driven convection in the water is governed by the mass balance of the dissolved gas and the mass and momentum balance of the water itself. The concentration of dissolved gas in the water, $C_w^*$, evolves due to both diffusive and convective transport following 
\begin{subequations}\label{eq:gov_pde}
\begin{eqnarray}
\dfrac{\partial{C_w^*}}{\partial t^*} + \nabla^*\cdot\left(\mathbf{u}_w^*C_w^*\right) = D{\nabla^*}^2{C_w^*}, \label{Solute}
\end{eqnarray}
where $D$ is the diffusivity and $\mathbf{u}_w^* = (u^*,w^*)$ the volume-averaged pore velocity.  The latter is given by Darcy's law and continuity, so that
\begin{eqnarray}
	&\mathbf{u}_w^* = - \dfrac{\perm}{\mu\varphi} \left(\nabla^*{P}_w^* + \rho_w^*g{\bf e}_{z^*}\right),\label{Darcy}\\
	&\nabla^*\cdot\mathbf{u}_w^* = 0,\label{Continuity}
\end{eqnarray}
\end{subequations}
where $\perm$ is the medium permeability, $\mu$ is the dynamic viscosity of the fluid, $\varphi$ is the porosity, $g$ is the acceleration of gravity and ${\bf e}_{z^*}$ is a unit vector in the $z^*$ direction. The density, $\rho_w^*$, is assumed to be a linear function of the concentration
\begin{eqnarray}
	\rho_w^* =  \rho^*_0 + \Delta\rho^*_0\dfrac{C^*_w}{C_{s, 0}^*},\label{Rho}
\end{eqnarray}
where $\rho_0^*$ is density of the fresh water and $\Delta \rho^*_0$ is the density difference between the fresh water and the saturated water at the initial pressure. The water contains no dissolved gas so that the initial condition is
\begin{eqnarray}
	\left.C_w^*\right|_{t^*=0}  = 0\;\; \mbox{for} \;\; z^* < 0. 
	\label{IC_pde}
\end{eqnarray}
The domain is $W$-periodic in the $x^*$ direction and impermeable to flow at top and bottom. At the bottom of the domain, the boundary conditions are homogeneous
\begin{subequations}\label{eq:BC_pde}
\begin{align}
	\left.\dfrac{\partial C^*_{w}}{\partial z^*}\right|_{z^* = -H_w} =0 \quad \mathrm{and}\quad  \left.w^*\right|_{z^* = -H_w} = 0. \label{eq:BC_bot}
\end{align}
The dissolved concentration at the interface is determined by local equilibrium with the gas, so that the boundary conditions at the top are given by
\begin{align}
	 \left.C_w^*\right|_{z^* = 0} =  C_{s}^*(t^*) \;\; \mbox{and} \;\; \left.w^*\right|_{z^* = 0} = 0. \label{eq:BC_top}
\end{align}
\end{subequations}
The evolution of the dissolved gas concentration at the interface, $C_s^*$, is determined by mass balance of the gas, given by
\begin{eqnarray}
	\frac{d n_g}{d t^*} = -AF^*, \label{GasMole}
\end{eqnarray}
where $A$ is the area of the interface at $z^* = 0$ (in the 2D system, $A = W$) and $F^*$ is the  mole flux from the gas into the water. This flux can be evaluated as
\begin{eqnarray}
	F^* =  \left.D\frac{\partial \overline{C_w^*}}{\partial z^*}\right|_{z^* = 0} = \frac{D}{W}\int_0^W\left.\frac{\partial C_w^*}{\partial z^*}\right|_{z^* = 0}dx^*, \label{Flux}
\end{eqnarray}
where the overline denotes the horizontal average as defined above. Combining (\ref{IdealGas}), (\ref{Henry}), (\ref{GasMole}), with (\ref{Flux}) results in the ODE for the evolution of the dissolved concentration at the interface
\begin{eqnarray}
	\frac{d C_s^*}{d t^*} = -\frac{K_hRT}{H_g}F^* = -\frac{K_hRTD}{H_g}\left.\frac{\partial \overline{C_w^*}}{\partial z^*}\right|_{z^* = 0}, \label{eq:gov_ode}
\end{eqnarray}
with the initial condition
\begin{align}
    \left.C_s^*\right|_{t^*=0}= C_{s, 0}^*=K_hP_{g,0}^*. \label{eq:ode_ic}
\end{align}
 This ODE is coupled to the system (\ref{eq:gov_pde}) through the flux (\ref{Flux}).

\subsection{Dimensionless equations}\label{sec:Dimenless_Eqns}
A uniform non-dimensionalization of the model problem is difficult, since the dominant length scales change with time \citep{Riaz2006,Hewitt2013shutdown,Slim2013}. Porous media convection is governed by the Rayleigh-Darcy number, $Ra=\U \H/D$, where $\U$ and $\H$ are suitable velocity and length scales, respectively \citep{Horton1945,Lapwood1948}. The Rayleigh-Darcy number is effectively a P\'{e}clet number and can be interpreted as the ratio between diffusive, $\Td = \H^2/D$, and advective, $\Ta = \H/\U$, timescales, $Ra = \Td/\Ta$.

The natural velocity scale in the convecting system is the buoyancy velocity, $\U = \perm\Delta\rho_0^* g/(\mu\varphi)$. Convection initiates along the top boundary and penetrates into the domain at a speed proportional to $\U$. At early time, after onset of convection but before convection spans the entire domain, the thickness of the diffusive boundary layer, $D/\U$, provides an natural length scale \citep{Riaz2006}. At later time, convection interacts with the bottom boundary and the domain height, $H_w$, is the appropriate length scale. Advection and diffusion balance across the boundary layer, so that the advective and diffusive timescales are identical at early time, $\Tad = D/\U^2$ \citep{Slim2014}. Scaling the system by the thickness of the diffusive boundary layer sets the Rayleigh number to unity and highlights the universal behaviour of the early convecting system. 

Below we assume that $\Td$ and $\Ta$ are based on $H_w$, appropriate for the long-term evolution of the convecting system. These two late time scales are related to the early time scale as follows
\begin{align}
\Tad = \Ta/Ra_0 = \Td/Ra_0^2,\quad\mathrm{where}\quad Ra_0 = \frac{\perm\Delta\rho_0^* g H_w}{\varphi\mu D},\label{eq:Ra}
\end{align}
is the Rayleigh-Darcy number based on the initial density difference. In a convecting system $Ra_0\gg 1$ so that the magnitudes of these timescales differ significantly.

To allow reduction of the governing equations to a purely diffusive system we choose the diffusive time, $\Td=H_w^2/D$, as characteristic timescale and define the following dimensionless variables
\begin{eqnarray}
\mathbf{x} = \dfrac{\mathbf{x}^*}{H_w},\;\;\;\rho = \dfrac{\rho_w^*}{\Delta \rho_0^*},\;\;\;t = \dfrac{t^*}{\mathcal{T}_d},\;\;\;\mathbf{u} = \dfrac{\mathbf{u}_w^*}{\U},\;\;\;\tilde{P} = \dfrac{P_w^*}{\Delta\rho_0^*gH_w},\;\;\;C = \dfrac{C_w^*}{C_{s, 0}^*}. \label{Nondim}
\end{eqnarray}
Substituting these scales into (\ref{eq:gov_pde}) leads to the following dimensionless governing equations
\begin{subequations}\label{eq:gov_pde_non_dim}
\begin{eqnarray}
& \dfrac{\partial C}{\partial t} + Ra_0\mathbf{u}\cdot\nabla C = {\nabla}^2 C, \label{Solute_nondim}\\
& \mathbf{u} = -{\nabla}P - C{\bf e}_{z}, \label{Darcy_nondim} \\
&\nabla\cdot\mathbf{u} = 0, \label{Continuity_nondim}
\end{eqnarray}
\end{subequations}
where $P = \tilde{P} + (\rho_0^*/\Delta\rho_0^*)z$ and $Ra_0$ is the initial Rayleigh-Darcy number defined in (\ref{eq:Ra}). This system of equations is solved subject to the following dimensionless initial condition
\begin{eqnarray}
	C|_{t=0}  = 0\;\; \mbox{for} \;\; z < 0, 
	\label{ICs_nondim}
\end{eqnarray}
and boundary conditions 
\begin{eqnarray}
	\left.\dfrac{\partial C}{\partial z}\right|_{z=-1} =  \left.w\right|_{z=-1} = 0; \;\; \quad \left.C\right|_{z=0} =  C_s(t) \;\; \mbox{and} \;\; \left.w\right|_{z=0} = 0, \label{BCs_nondim1}
\end{eqnarray}
where $C_s$ is the dimensionless dissolved concentration at the interface. Note that here $C_s$ is also identical to the normalised gas pressure from Henry's law and ideal gas law, i.e. $C_s = P_g^*/P_{g,0}^*$. The evolution of $C_s$ is given by the following ODE and initial condition
\begin{align}
    \dfrac{d C_s}{d t} = -\param\left.\frac{\partial \overline{C}}{\partial z}\right|_{z = 0}\quad\mathrm{and} \quad C_s|_{t=0}  = 1, \label{eq:ODE_non_dim}
\end{align}
where $\param$ is the dissolution capacity, defined by (\ref{eq:new_param}). {\color{black}{The equation (\ref{eq:ODE_non_dim}) actually works as a Robin boundary condition for the concentration field in the water.  Similar boundary conditions are also utilised in some thermal porous media convection with imperfectly conducting boundaries \citep{Wilkes1995,Kubitschek2003,Barletta2012,Barletta2015,Hitchen2016}, where the heat flux depends linearly on the surface temperature and a dimensionless parameter $Bi$, the Biot number, is characterised to represent the rate of thermal transport across the boundary. However, unlike those thermal convection studies, here the \emph{time-dependent} equation (\ref{eq:ODE_non_dim}) couples a global mass balance between two subsystems (i.e. the gas and the water) and is always uniform along the gas-water interface.}} 

The dimensionless dissolution flux $F$ that couples (\ref{eq:gov_pde_non_dim}) and (\ref{eq:ODE_non_dim}) can be expressed as 
\begin{eqnarray}
	F(t) =  \left.\dfrac{\partial \overline{C}}{\partial z}\right\vert_{z=0} = \dfrac{1}{L}\int^{L}_0\left.\dfrac{\partial C}{\partial z}\right\vert_{z=0}dx. \label{Flux_nondim}
\end{eqnarray}
Comparing (\ref{Flux}) and (\ref{Flux_nondim}), the scale for the flux is $\mathcal{F} =DC_{s,0}^*/H_w$, so that  $F = F^*/\mathcal{F}$.  To measure the magnitude of the CO$_2$ dissolution, we define the volume-averaged concentration in the water 
\begin{eqnarray}
	\overline{\overline{C}}(t) =  \dfrac{\mathbf{\int} Cd\mathbf{x}}{V_w},  \label{C_volavg}
\end{eqnarray}
and mass conservation of the whole system requires that
\begin{eqnarray}
	C_s + \param\overline{\overline{C}} \equiv  1.  \label{Cs_C_volavg}
\end{eqnarray}

While the governing equations have been scaled by the diffusion time $\Td$, other scales may be appropriate for the discussion of early and late phenomena. Therefore, we define the following diffusive, advective (or convective), and advective-diffusive dimensionless times
\begin{align}
    t = t_d = t^*/\Td,\quad t_a = t^*/\Ta, \quad \mathrm{and} \quad t_{ad} = t^*/\Tad,
\end{align}
respectively.


\subsection{Numerical method}\label{sec:Numerical}
To solve these governing equations numerically, it is convenient to first introduce a stream function $\psi$ to describe the 2D fluid velocity, so that $\mathbf{u} = (u,w) = (\partial_z\psi, -\partial_x\psi)$ and the continuity equation (\ref{Continuity_nondim}) is satisfied.  Then the dimensionless equations (\ref{Darcy_nondim}) and (\ref{Solute_nondim}) can be written as
\begin{eqnarray}
	& \nabla^2{\psi} = \partial_x C, \label{Psi}\\
	& \partial_tC + Ra_0(\partial_z\psi\partial_xC - \partial_x\psi\partial_zC) = \nabla^2 C, \label{Solute_Psi}
\end{eqnarray}
where $\psi$ satisfies $L$-periodic boundary conditions in $x$ and homogeneous Dirichlet boundary conditions in $z$.

In our study, the equations (\ref{Psi}) and (\ref{Solute_Psi}) were solved numerically using a Fourier--Chebyshev-tau pseudospectral algorithm \citep{Boyd2000}.  For temporal discretization, a third-order-accurate semi-implicit Runge--Kutta scheme \citep{Nikitin2006} was utilised for computations of the first three steps, and then a four-step fourth-order-accurate semi-implicit Adams--Bashforth/Backward--Differentiation scheme \citep{Peyret2002} was used for computation of the remaining steps. 
At each step, we updated $C_s$ by solving (\ref{eq:gov_ode})  explicitly using a two-step Adams--Bashforth algorithm.  

\section{Diffusion solution and onset of convection}\label{sec:Diffusion_onset}
At sufficiently small Rayleigh number, the system is stable to perturbations and mass transfer is purely diffusive. When the Rayleigh number is above some critical value, the diffusive boundary layer becomes unstable and induces downward moving convective fingers which significantly increase the rate of CO$_2$ dissolution into the water.  In \S\,\ref{sec:Diffusion_Solution} we provide analytic approximations for the diffusive base state in the closed system   and then study the onset of the convection using DNS in \S\,\ref{sec:Onset}.

\subsection{Diffusion solution}\label{sec:Diffusion_Solution}
In a stable system $\mathbf{u}=0$ and mass transport is purely diffusive, so that (\ref{eq:gov_pde_non_dim}) reduces to the one-dimensional diffusion equation 
\begin{eqnarray}
	\dfrac{\partial C}{\partial t}  = \dfrac{\partial^2 C}{\partial z^2} \quad\mathrm{on}\quad z\in\left[-1,\, 0\right], \label{Diffusion}
\end{eqnarray}
with (\ref{ICs_nondim}) and (\ref{BCs_nondim1}) as  initial and boundary conditions, respectively. A complete closed form solution is not available, but solutions in different limiting cases can be obtained by using {\color{black}{a}} Laplace transform. For $\param = 0$, the classic series solution for diffusion in a finite domain can be written as
\begin{eqnarray}
	C(z,t)  =  \sum_{n = 0}^{\infty} (-1)^n \left[ 2 - \mbox{erf}\left(\dfrac{-z}{2\sqrt{t}} +  \dfrac{n}{\sqrt{t}}\right) 
	-  \mbox{erf}\left(\dfrac{z}{2\sqrt{t}} + \dfrac{n+1}{\sqrt{t}}\right)\right],	\label{finite_eta0}
\end{eqnarray}
on $z\in\left[-1,\, 0\right]$ \citep{Kim2015}. This solution reduces to simple error function solution for diffusion in a semi-infinite domain at early time, $t \ll 1$.
For $\param>0$, a closed form solution can only be found at early time when the domain is effectively semi-infinite. This solution is then given by
\begin{eqnarray}
	C(z,t)  = e^{\param^2t - \param z}\left[1 + \mbox{erf}\left(-\param\sqrt{t} + \dfrac{z}{2\sqrt{t}}\right)\right]  \quad\mathrm{on}\quad z\in\left(-\infty,\, 0\right],	 \label{Soln_semi-infinite}
\end{eqnarray}
and reduces to the standard error function solution in the limit $\param=0$. Hereafter, (\ref{Soln_semi-infinite}) is referred to as the \emph{early-time} solution. We note that these solutions are not self-similar in $z/\sqrt{t}$,  if $\param > 0$.

At late time, the diffusive front interacts with the bottom boundary and {\color{black}{the finiteness of the domain affects the solution}}. For $\param>0$, the full solution in the Laplace transform variable is given by
\begin{eqnarray}
\lpls{C}(z,s)= \frac{\cosh(\sqrt{s}(z+1))}{s\cosh(\sqrt{s})+\param\sqrt{s}\sinh(\sqrt{s})} \quad\mathrm{on}\quad z\in\left[-1,\, 0\right], \label{Soln_laplace-domain}
\end{eqnarray}
but the inverse Laplace-transform of this expression does not lead to a closed form expression. Instead, a series solutions can be obtained via  Cauchy's residue theorem  \citep{Duffy2004}.  This requires the poles, $s_k$, of (\ref{Soln_laplace-domain}), which are given implicitly by the roots of
\begin{eqnarray}
\tan(p) = -\frac{p}{\param}, \label{Soln_Poles2}
\end{eqnarray}
where $s=-p^2$ \citep{Zhang2017}. From the definition of the inverse Laplace transform and Jordan's Lemma \citep{Schiff1999}, the solution is then given by 
\begin{eqnarray}
C(z,t) = \mathcal{L}^{-1}\left \{  \lpls{C}(z,s) \right \} = \frac{1}{2\pi i} \lim_{\widetilde{T}\to\infty} \int_{r-i\widetilde{T}}^{r+i\widetilde{T}} \lpls{C}(z,s)e^{st} ds = \sum_{n=0}^{\infty} a_{n}e^{-p_{n}^{2}t}, \label{Cauchy2}
\end{eqnarray}
where the coefficients of the residues for the simple poles are
\begin{eqnarray}
  a_n = \lim_{s\to s_n}  (s-s_n)\lpls{C}(z,s) = \left\{
    \begin{array}{ll}
     \quad \quad \quad \dfrac{1}{1+\param} & n =0, \\ \\
     \dfrac{2\param \cos(p_{n}z)+2p_{n}\sin(p_{n}z)}{\param^2+\param+p_{n}^2}        & n \ge 1.
    \end{array} \right. \label{Residues}
\end{eqnarray}
At late times (\ref{Cauchy2}) is dominated by lowest order terms and the equilibrium solution is given by the zeroth-order term 
\begin{eqnarray}
	\lim_{t \rightarrow \infty}C  = \dfrac{1}{1 + \param}.	 \label{Equilibrium}
\end{eqnarray}
The equilibrium solution is constant and entirely determined by the dissolution capacity, $\param$. At equilibrium, {\color{black}{$C=C_s= \overline{\overline{C}}$, so that (\ref{Equilibrium}) is consistent with the equilibrium condition from overall mass balance (\ref{eqbm}) and mass conservation (\ref{Cs_C_volavg}).}} The equilibrium concentration declines rapidly with increasing dissolution capacity, as a decreasing amount of gas dissolves into an increasing amount of water.

\begin{figure}[t]
{
  \centering
  \sidesubfloat[]{\includegraphics[height=1.85in]{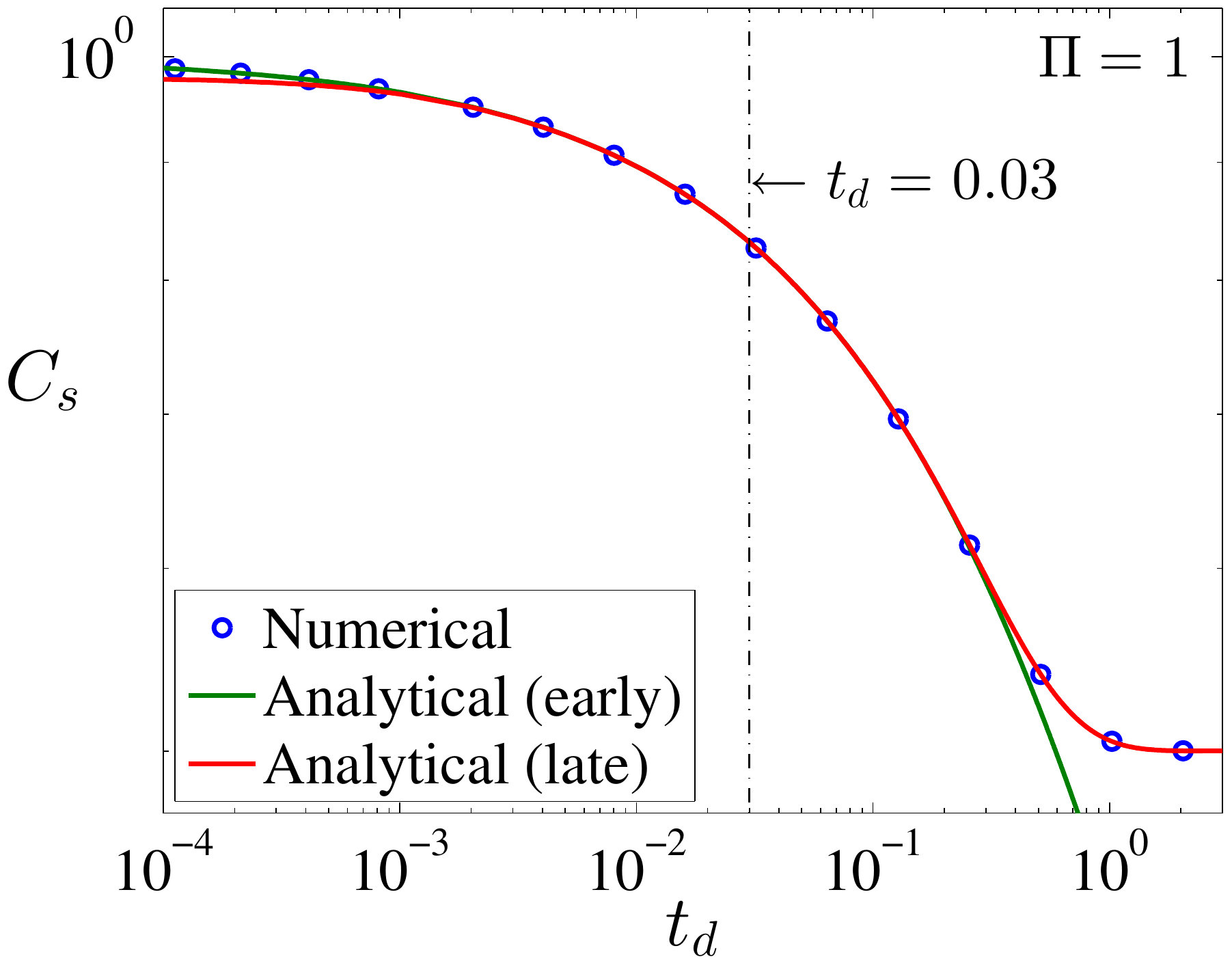}} \quad
  \sidesubfloat[]{\includegraphics[height=1.85in]{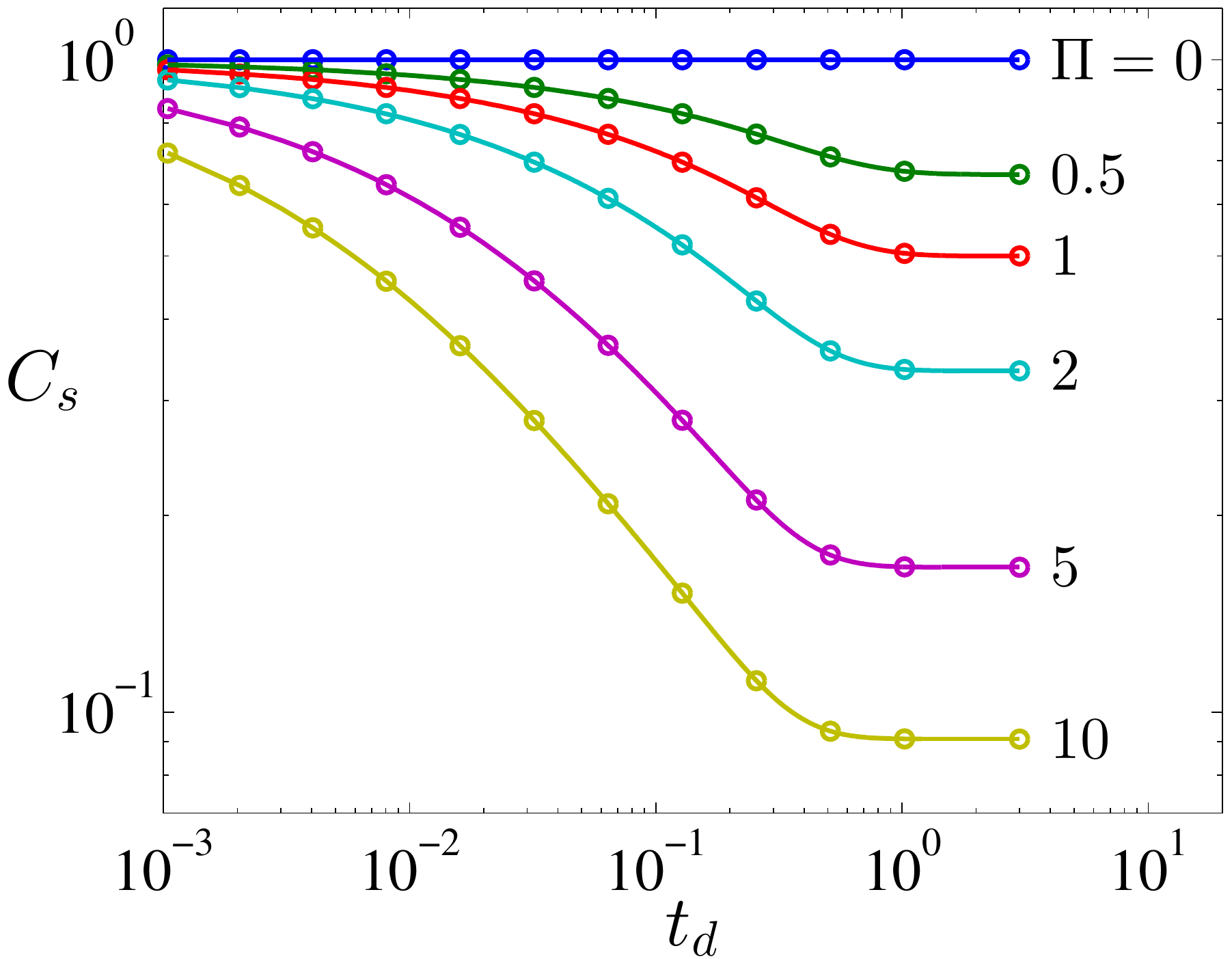}} \\
  \sidesubfloat[]{\includegraphics[height=1.85in]{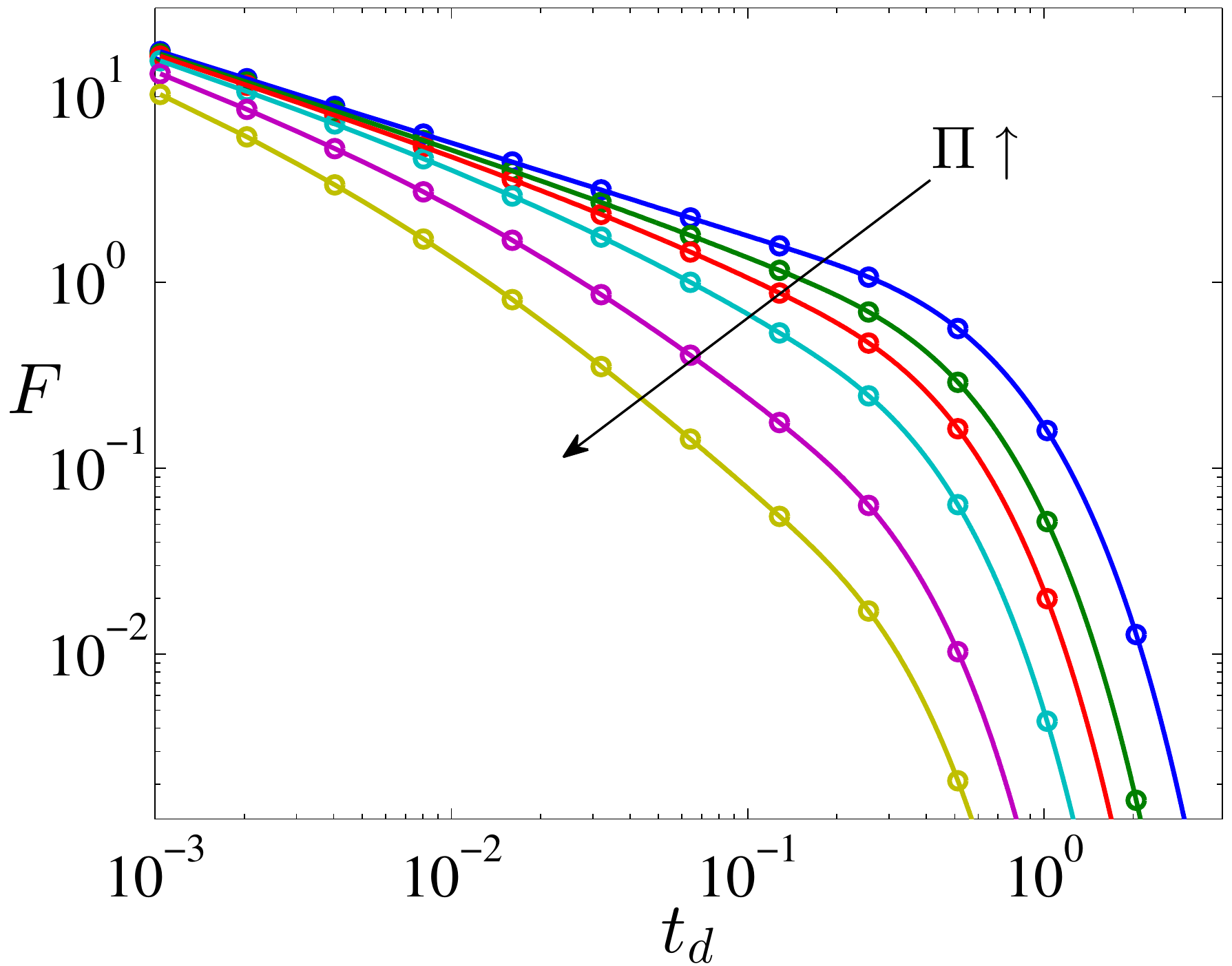}} \quad
  \sidesubfloat[]{\includegraphics[height=1.85in]{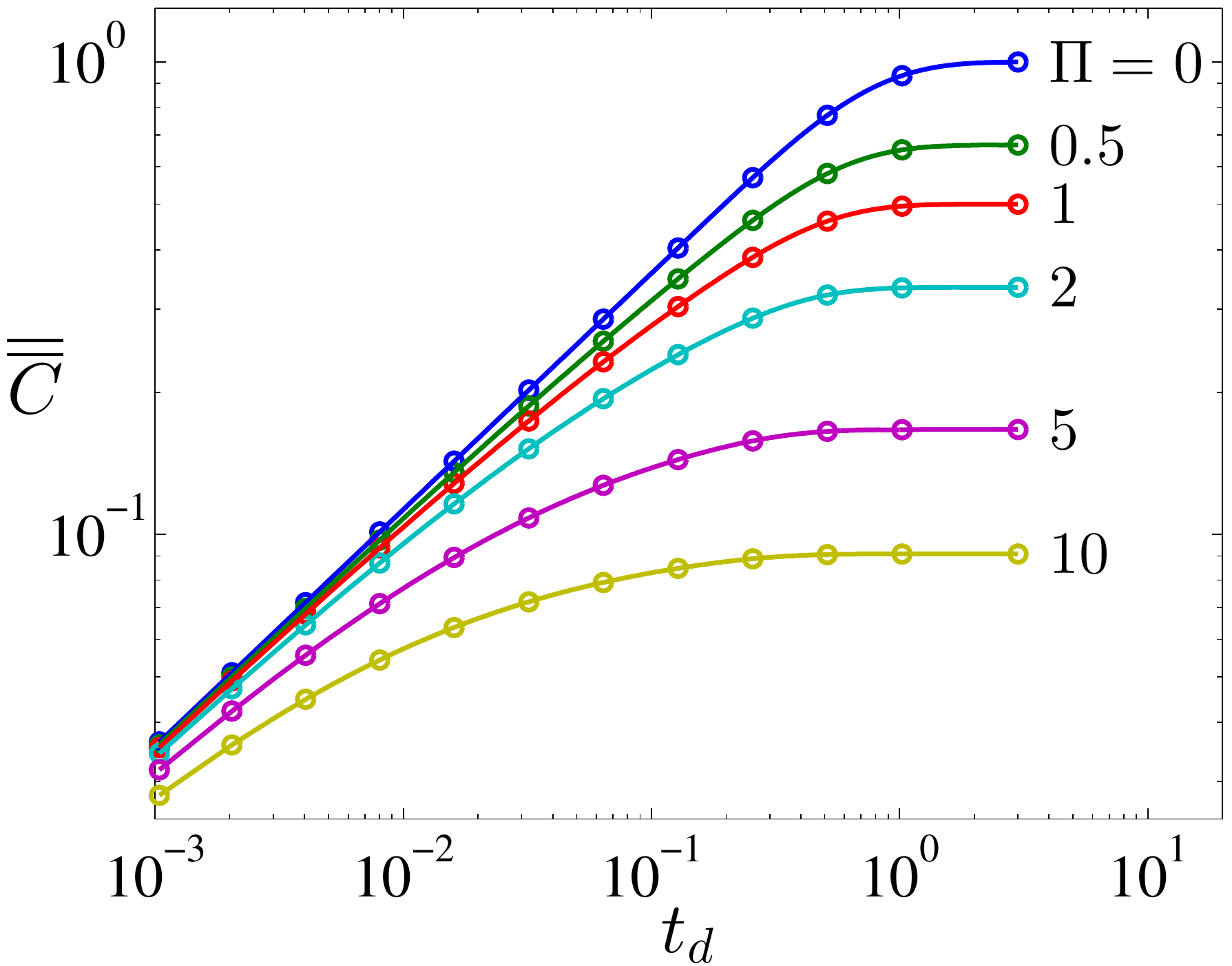}} \par
}
\caption{Comparisons between numerical and analytic diffusion solutions in the closed system. Circles: numerical solution (the spacing of symbols does not reflect the time steps used in computations); solid lines: analytic solution. In $(a)$, the solutions (\ref{Soln_semi-infinite}) and (\ref{Cauchy2}) are valid, respectively, in early and late times; in $(b)$--$(d)$ and for $\param > 0$, the early-time solution is used as the composite analytic solution at $t_d \le 0.03$ and the late-time one is used at $t_d > 0.03$. 10 modes (i.e. $0 \le n \le 9$) are utilised to calculate the analytic solutions (\ref{finite_eta0}) and (\ref{Cauchy2}). However, our study indicates that for $t_d \ge 0.1$, 2 modes (i.e. $n = 0$, 1) of the late-time solution (\ref{Cauchy2}) are enough to retain a 99\% accuracy of the quantities showing in $(b)$--$(d)$.}  \label{fig:DiffusionSoln}
\end{figure}

The low-order terms in (\ref{Cauchy2}) generally capture the late-time behaviour, but a large number of modes is needed to describe the solution at early time. We therefore truncate the sum in (\ref{Cauchy2}) to obtain a \emph{late-time} approximation and combine it with early-time solution, given by (\ref{Soln_semi-infinite}), to describe the full evolution.  Figure~\ref{fig:DiffusionSoln}($a$) shows this composite solution for the concentration on the interface, $C_s(t) = \left.C\right|_{z=0}$, and a numerical solution using the algorithms described in \S\,\ref{sec:Numerical} matches this composite analytic solution well. For this comparison, the numerical solution was initialised with (\ref{Soln_semi-infinite}) evaluated at $t = 4\times10^{-5}$ to avoid oscillations arising from the discontinuity between the initial and the boundary conditions.

\floatsetup[figure]{style=plain,subcapbesideposition=top}  

The early-time solution gives insight into the effect that $\param$ has on the diffusive mass transport in a closed system. The concentration on the interface and the flux across the interface are given by
\begin{subequations}
\begin{align}
    C_s(t) &= C(t,0) = e^{\param^2t}\left[1 + \mbox{erf}\left(-\param\sqrt{t}\right)\right],	 \label{Cs_early}\\
    F(t,0) &=  -\param e^{\param^2t}\left[1 + \mbox{erf}\left(-\param\sqrt{t}\right)\right] + \dfrac{1}{\sqrt{\pi t}},	 \label{F_early}
\end{align}
\end{subequations}
for $t\ll1$. In an open system $C_s$ is constant, but in closed systems $C_s$ declines ever more rapidly with increasing $\param$, as shown in figure~\ref{fig:DiffusionSoln}($b$). For $\param>0$, the period for which $C_s > 0.99$, i.e. approximately constant, is  $t < 0.009/\param$, so that the decline in $C_s$ begins earlier with increasing $\param$. In an open system, $F$ declines as $t^{-1/2}$ at $t\ll1$, but figure~\ref{fig:DiffusionSoln}($c$) shows that the flux in the closed system does not follow a simple power law, since the rapid decline of $C_s$ at early time reduces the diffusive flux much faster.

The negative feedback introduced by the mass balance constraint in a closed system significantly slows down both the rate of dissolution and the total amount that can be dissolved. However, the time required to reach global equilibrium, {{\color{black}{$t \approx 1$}}, remains approximately constant {{\color{black}{for different $\param$}}, as the reduction in flux is offset by the reduction in the equilibrium concentration (see figure~\ref{fig:DiffusionSoln}$d$). 

\subsection{Onset of convection}\label{sec:Onset}
The diffusive boundary layer will grow with time as the CO$_2$ continuously dissolves into the water.  When the diffusion layer becomes thick enough, the CO$_2$-rich water, which is heavier than the underlying fresh water, {\color{black}{can become}} unstable under the influence of gravity and {\color{black}{sink}} in plumes of  heavy CO$_2$-rich fluid.  This phenomenon, known as onset of convection, has been studied extensively by using linear stability analysis and DNS \citep{Ennis-King2005, Riaz2006, Xu2006,  Hassanzadeh2006, Kim2008, Kim2012, Slim2010, Pau2010, Javaheri2010, Elenius2012, Elenius2014,Tilton2014, Slim2014, Kim2015}.  A full hydrodynamic stability analysis for the closed system is beyond the scope of this contribution, but we provide DNS that illustrate the effect of $\param$ on the onset of convection. Simulations are conducted for a discrete set of $Ra_0$ and $\param$ in a 2D domain with aspect ratio $L = 10^5/Ra_0$. 

The concentration field $C(\mathbf{x},t)$ can be decomposed into a transient diffusive base state $C_d(z,t)$ plus a fluctuation $\widetilde{C}(\mathbf{x},t)$, namely, 
\begin{eqnarray}
	C(\mathbf{x},t) = C_d(z,t) +  \widetilde{C}(\mathbf{x},t),	 \label{Cdecomposition}
\end{eqnarray}
where the diffusion solution $C_d$ is a composite analytic solution as in figure~\ref{fig:DiffusionSoln} and the fluctuation term can be expressed as 
\begin{eqnarray}
	\widetilde{C}(\mathbf{x},t) = \sum_{n=-N/2}^{N/2}\hat{C}_n(z,t)e^{inkx},	 \label{Ctilde}
\end{eqnarray}
where $k = 2\pi/L$ is the fundamental wavenumber and $N$ is the horizontal truncation mode number.  In our DNS, the initial condition
is the the early-time solution at $t = 1/Ra_0^2$, corresponding to $t_{ad} = 1$, with random perturbations within the top diffusion layer. For the purpose of this study we define the onset of convection as the earliest time when the norm of the amplitude $\hat{C}_n$ starts to grow.


\begin{figure}[t]
{
  \centering
  \includegraphics[width=0.7\textwidth]{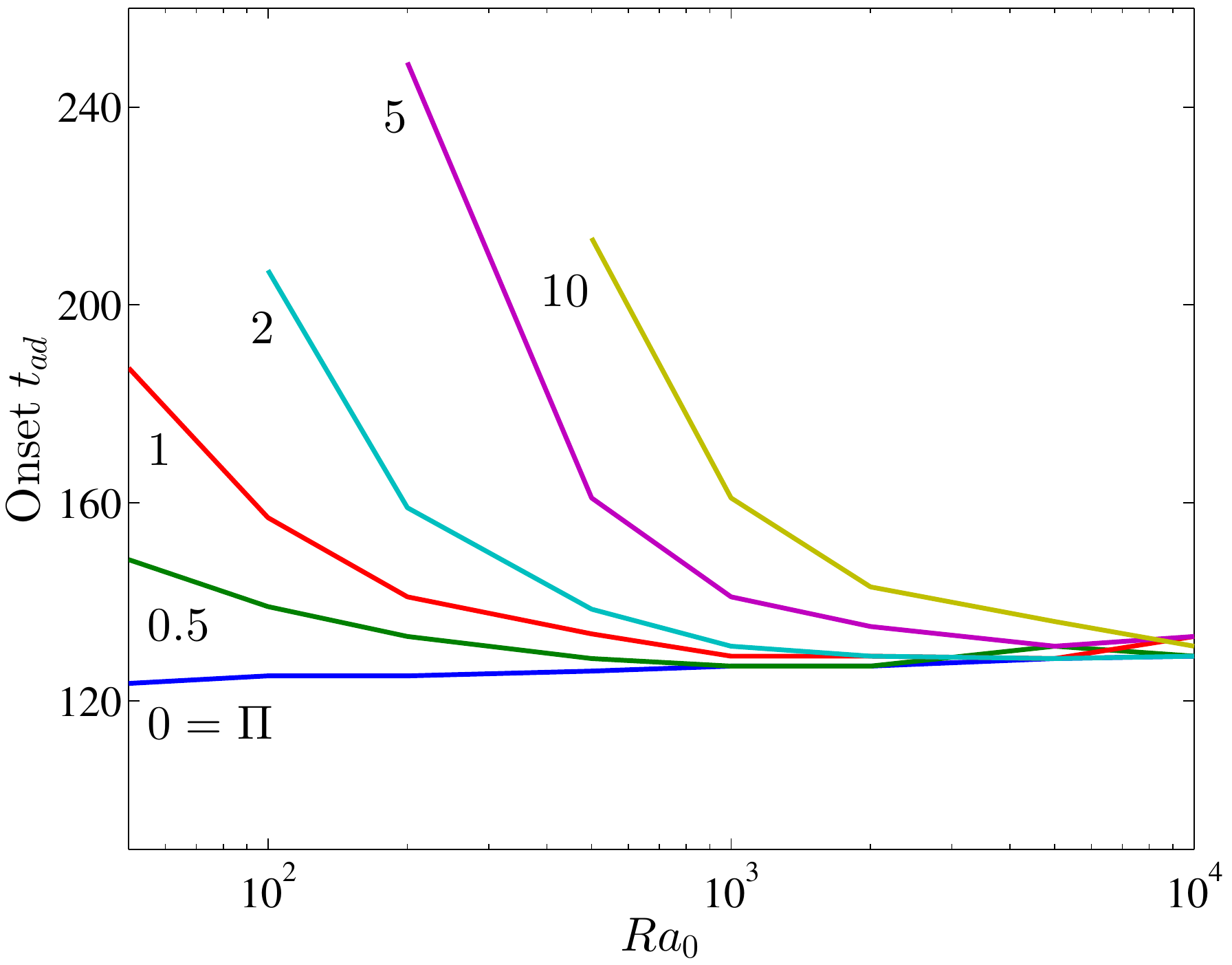}
}
\caption{Variation of onset time as a function of Rayleigh number and dissolution capacity. For $\param \ge 2$, The diffusion solution becomes stable at sufficiently small $Ra_0$ where instability has set in for $\param=0$. In closed systems, the onset of convection is not affected by $\param$ at large $Ra_0$, but delayed at small $Ra_0$ with increasing $\param$ due to the reduction of dissolution flux from the negative feedback of the pressure drop in gas.}  \label{fig:Onset}
\end{figure}

For $\param = 0$, the system of equations (\ref{eq:gov_pde_non_dim}) becomes parameter-less in the advective-diffusive scheme by rescaling $\mathbf{x} = \mathbf{x}_{ad}/Ra_0$ and $t = t_{ad}/Ra_0^2$, so that $Ra_0$ becomes the height of the rescaled layer and the solution is universal before the fingertips reach the bottom boundary.  As shown in figure~\ref{fig:Onset}, our DNS results indicate that the diffusion solution becomes unstable at $t_{ad} \approx 130$ for the open system. This is consistent with previous work on linear stability analysis, which gives $t_{ad} \approx 146$ \citep{Riaz2006, Javaheri2010, Elenius2014}. Nevertheless, for $\param > 0$, the system is $Ra_0$-dependent even in the advective-diffusive scaling as $Ra_0$ appears in (\ref{eq:ODE_non_dim}) {\color{black}{(for advective-diffusive scalings, the time $\Tad = \Td/Ra_0^2$ and the length $\H_{ad} = H_w/Ra_0$)}}. The onset time in a closed system therefore depends on both $Ra_0$ and $\param$ (see figure~\ref{fig:Onset}). For small $Ra_0$ the diffusive boundary layer has to grow to a larger thickness before instability occurs. This allows the negative feedback in a closed system to reduce the diffusive flux and to increase the onset time with increasing $\param$ (see figures~\ref{fig:DiffusionSoln}$c$ and \ref{fig:Onset}). At sufficiently large $Ra_0$, however, the onset occurs before the negative feedback in a closed system has reduced the diffusive flux, so that the onset time is independent of $\param$.




\section{Numerical simulations and mathematical models {\color{black}{at large $Ra$}}}\label{sec:DNS_model}
To study the dynamics and mass transport of solutal convection in the closed porous media system, DNS were performed at $Ra_0 = 20000$ for $\param = 0$, 0.5, 1, 2, 5 and 10 in a 2D domain with the aspect ratio $L = 10^5/Ra_0$.  In these computations, 8192 Fourier modes were utilised in the lateral discretization, 385 Chebyshev modes were used in the vertical discretization, and the time step is $\Delta t = 10^{-8}$.  Moreover, the early-time solution for the diffusive base state, given by (\ref{Soln_semi-infinite}), at time $t = 25/Ra^{2}_0$ (or $t_{ad} = 25$) was used as the initial condition for the concentration field, and a small random perturbation was added as a noise within the upper diffusive boundary layer to induce the convective instability.  Although the results only from $Ra_0=20000$ were utilised for following analysis, {\color{black}{it will be shown at the end of section~\ref{sec:models} our mathematical models}} are also applicable to other large Rayleigh numbers.

\subsection{DNS results}\label{sec:DNSresults}

\floatsetup[figure]{style=plain,subcapbesideposition=top}  
\begin{figure}[t]
{
  \centering
  \sidesubfloat[]{\includegraphics[height=1.9in]{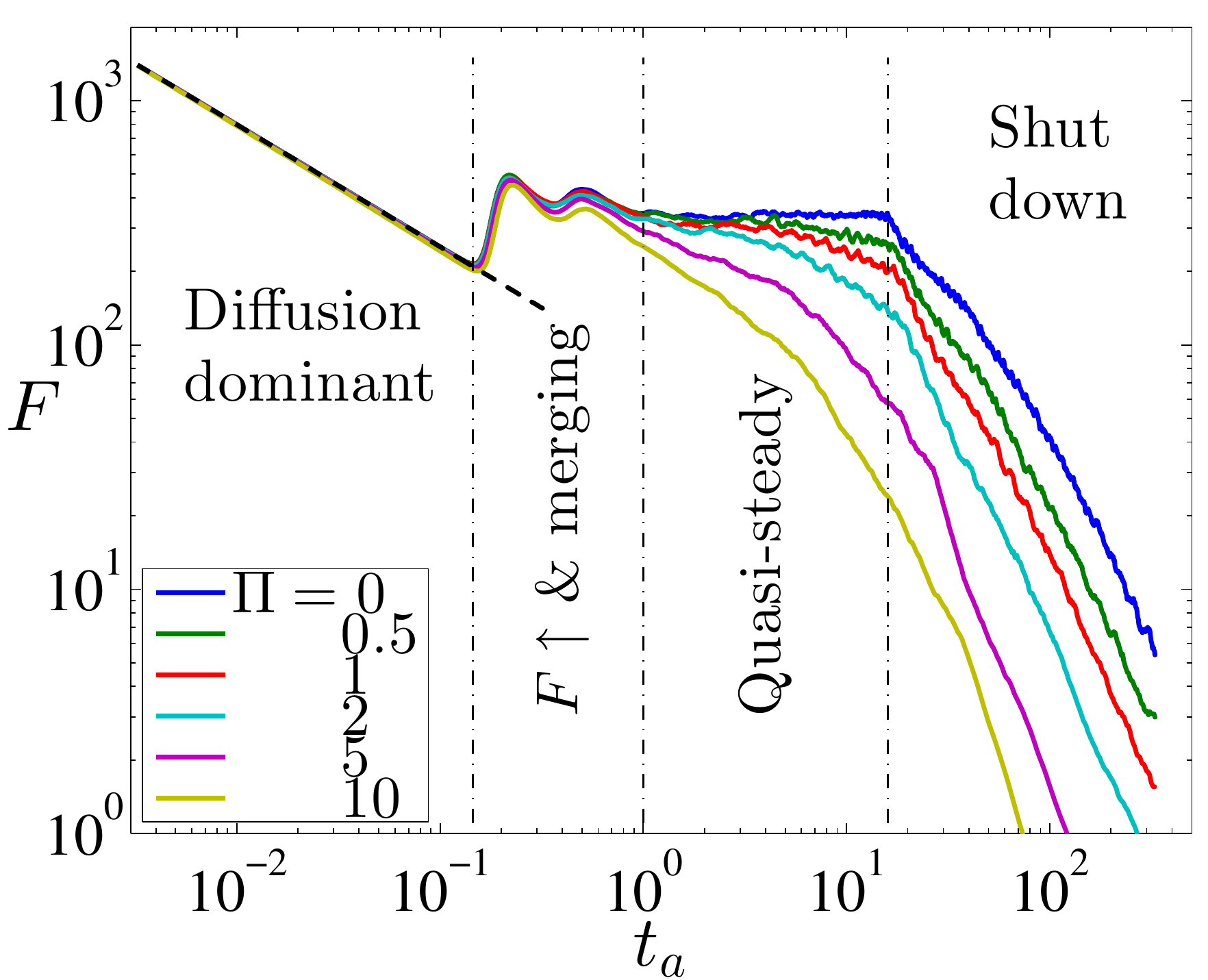}} \quad
  \sidesubfloat[]{\includegraphics[height=1.9in]{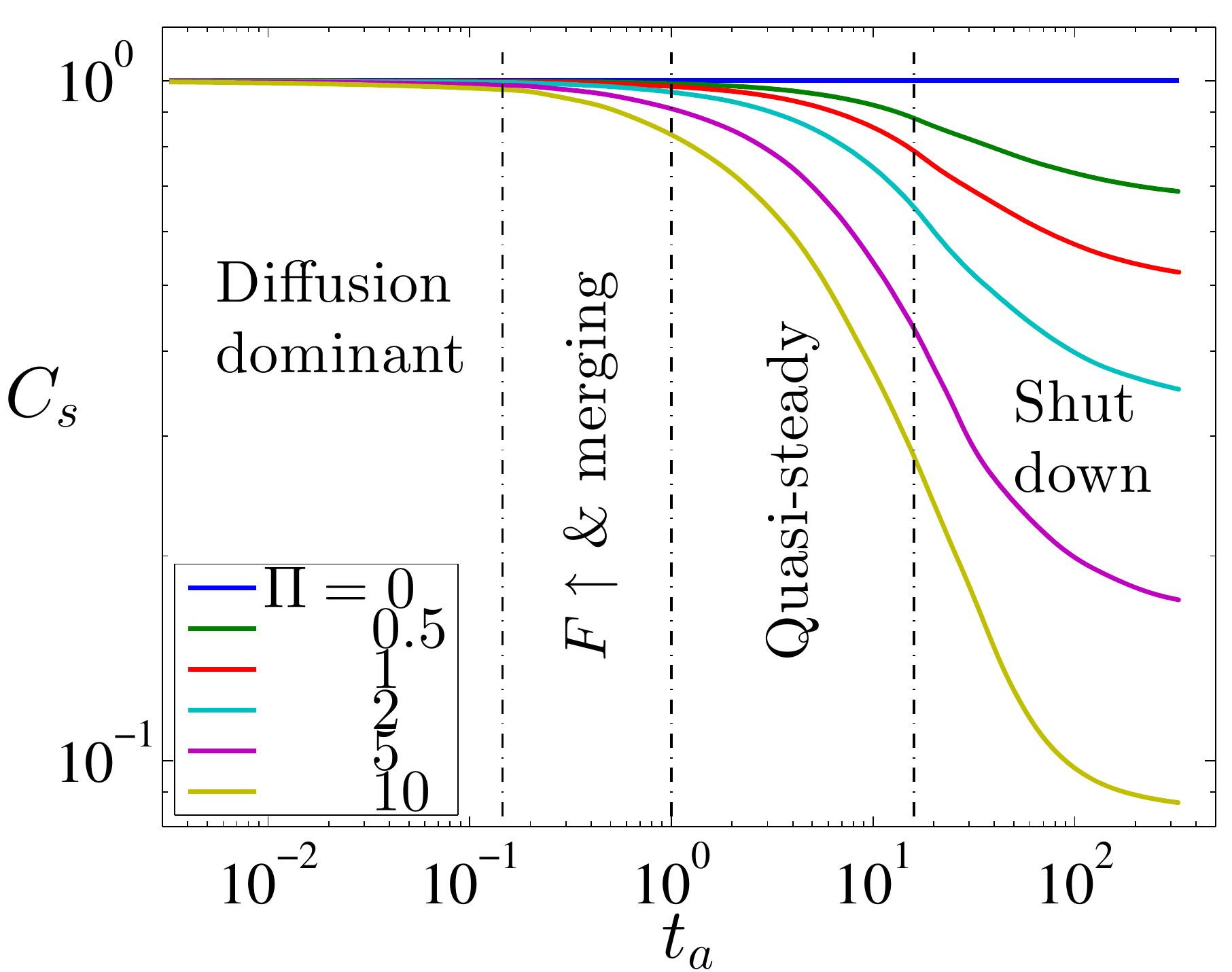}}\par
}
\caption{Evolution of the dissolution flux $F$ and the interface concentration $C_s$ in time at $Ra_0 = 20000$ for different $\param$.  Solid lines: DNS results; dashed lines in ($a$): purely diffusive flux $F\sim(\pi t)^{-1/2}$ for $\param=0$.  In both ($a$) and ($b$), the four dynamical regimes are delineated using dashed-dot lines based on the flow characteristics at $\param=0$.  Generally, these four regimes still exist for $\param>0$; however, for various $\param$ the time of transitions may be different, e.g. as analysed below the onset of shut-down regime will be delayed at a higher $\param$.  At sufficiently large $Ra_0$, $\param$ starts to affect the convection in the flux-growth \& plume-merging regime; and in the quasi-steady convective regime, the flux for $\param>0$ does not remain constant due to the decay of the interface concentration $C_s$.}  \label{fig:DNS_Cs_Flux_Ra20000}
\end{figure}

\floatsetup[figure]{style=plain,subcapbesideposition=top}  
\begin{figure}[t]
{
  \centering
  \sidesubfloat[]{\includegraphics[height=2.2in]{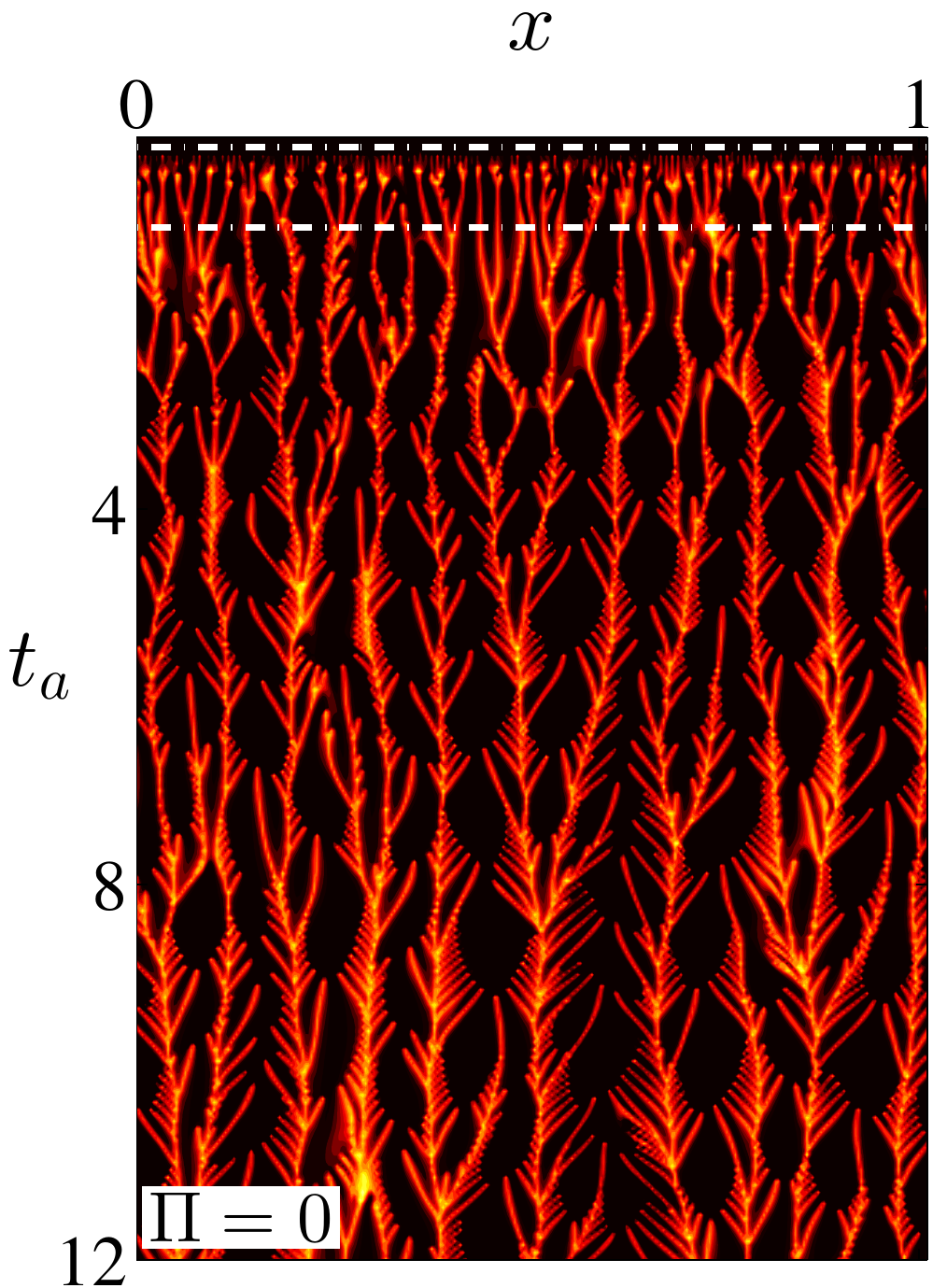}	
                         \includegraphics[height=2.2in]{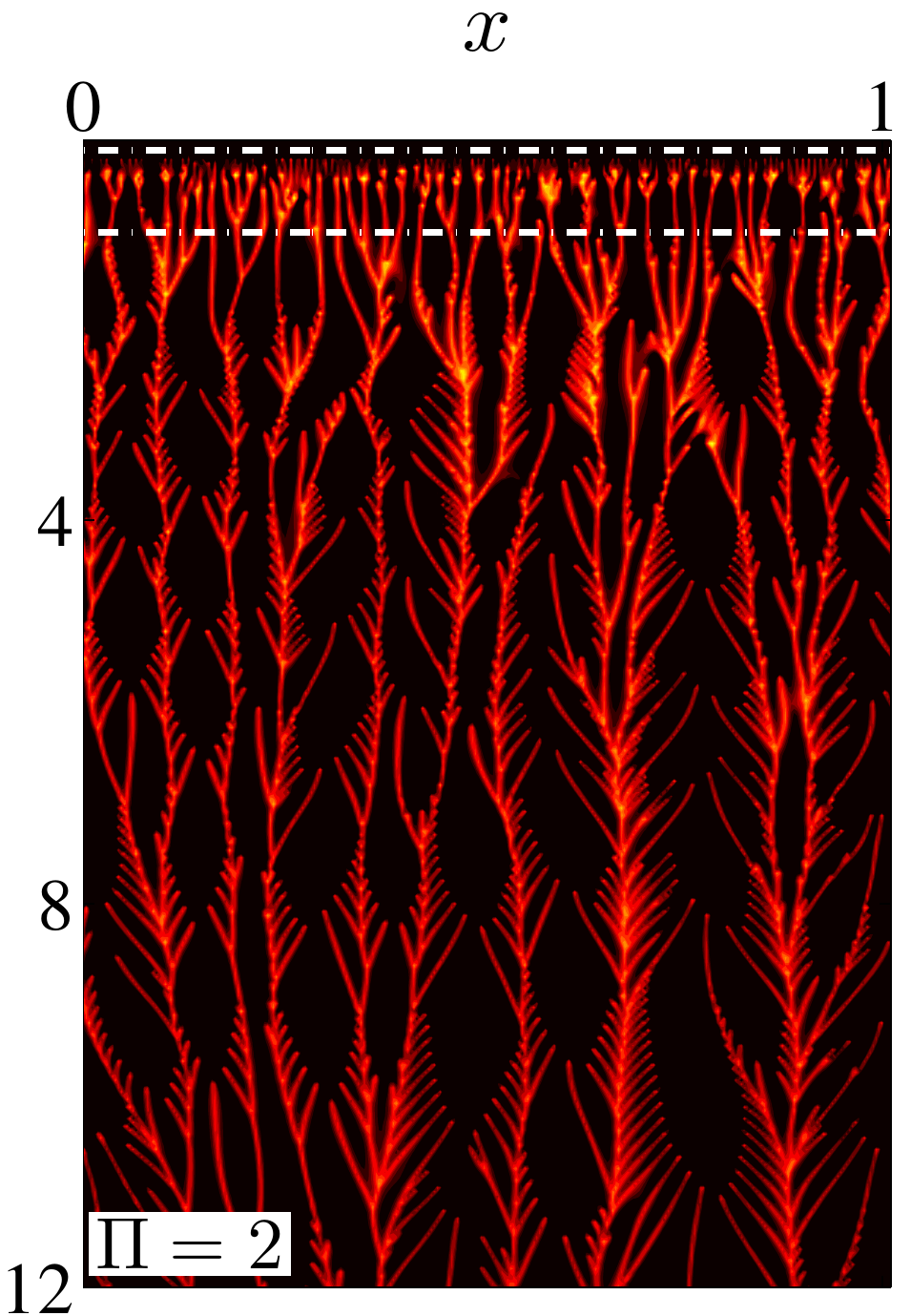}
                         \includegraphics[height=2.2in]{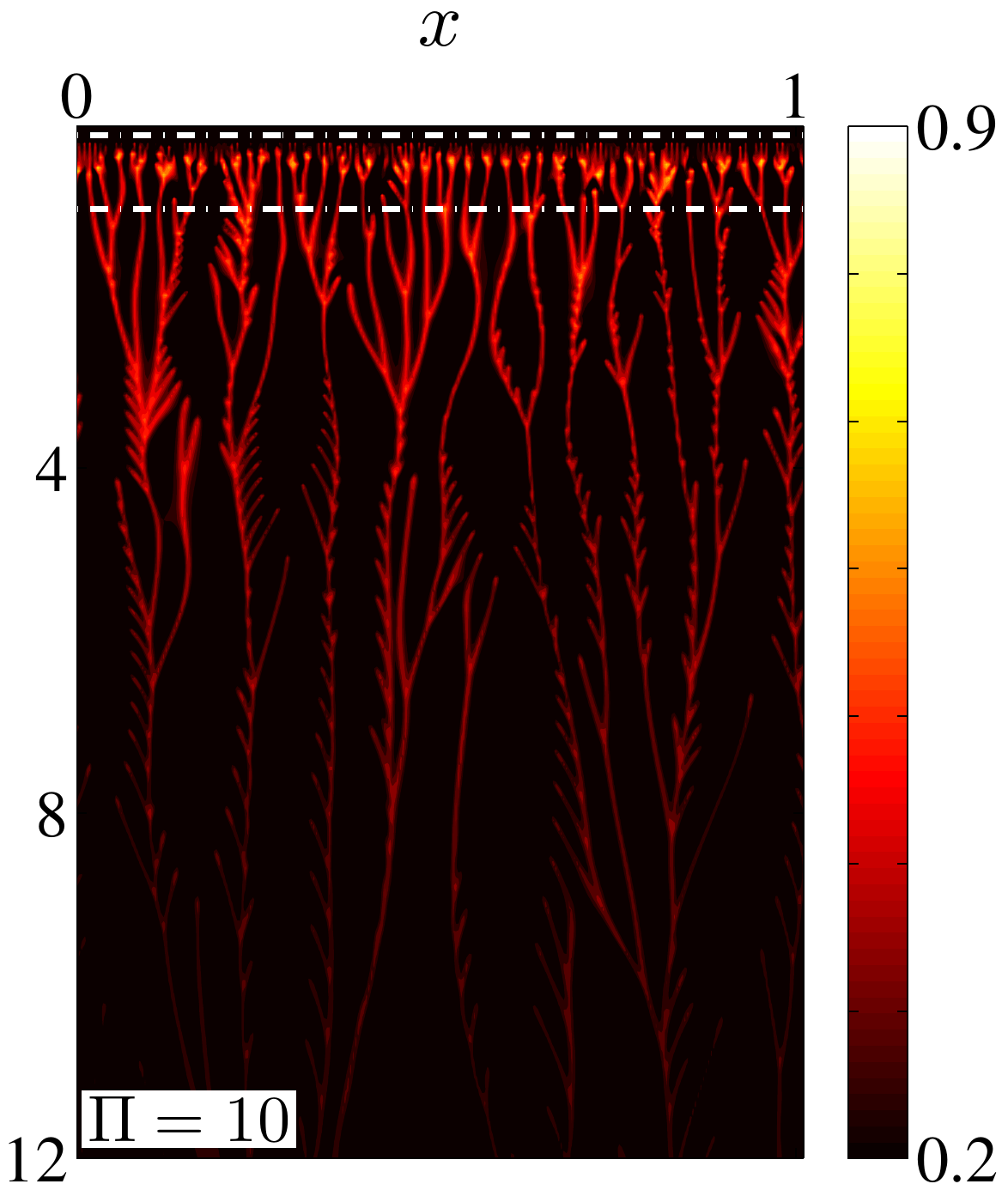}} \\\vspace{0.05in}
  \sidesubfloat[]{\includegraphics[height=1.9in]{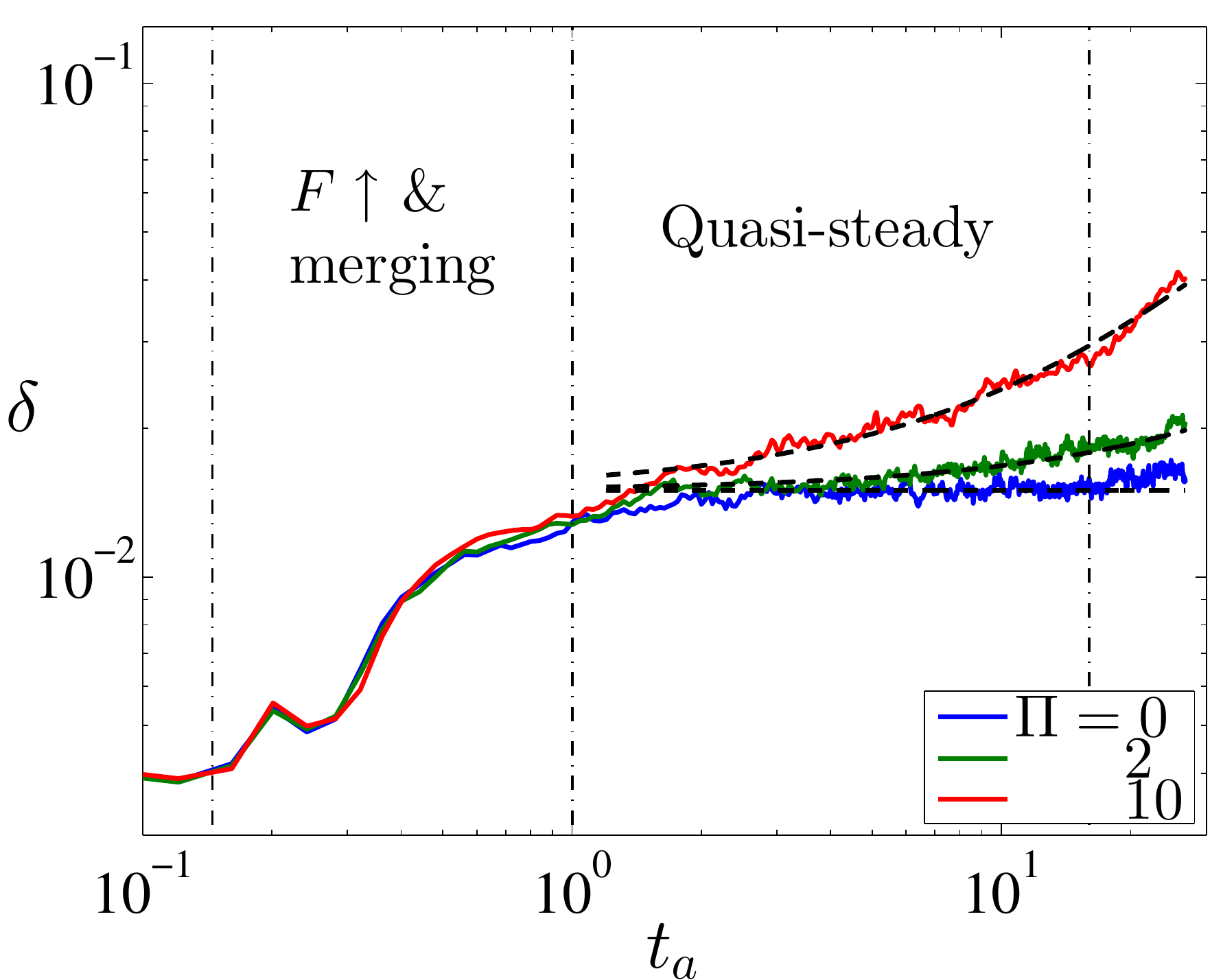}} \quad
  \sidesubfloat[]{\includegraphics[height=1.9in]{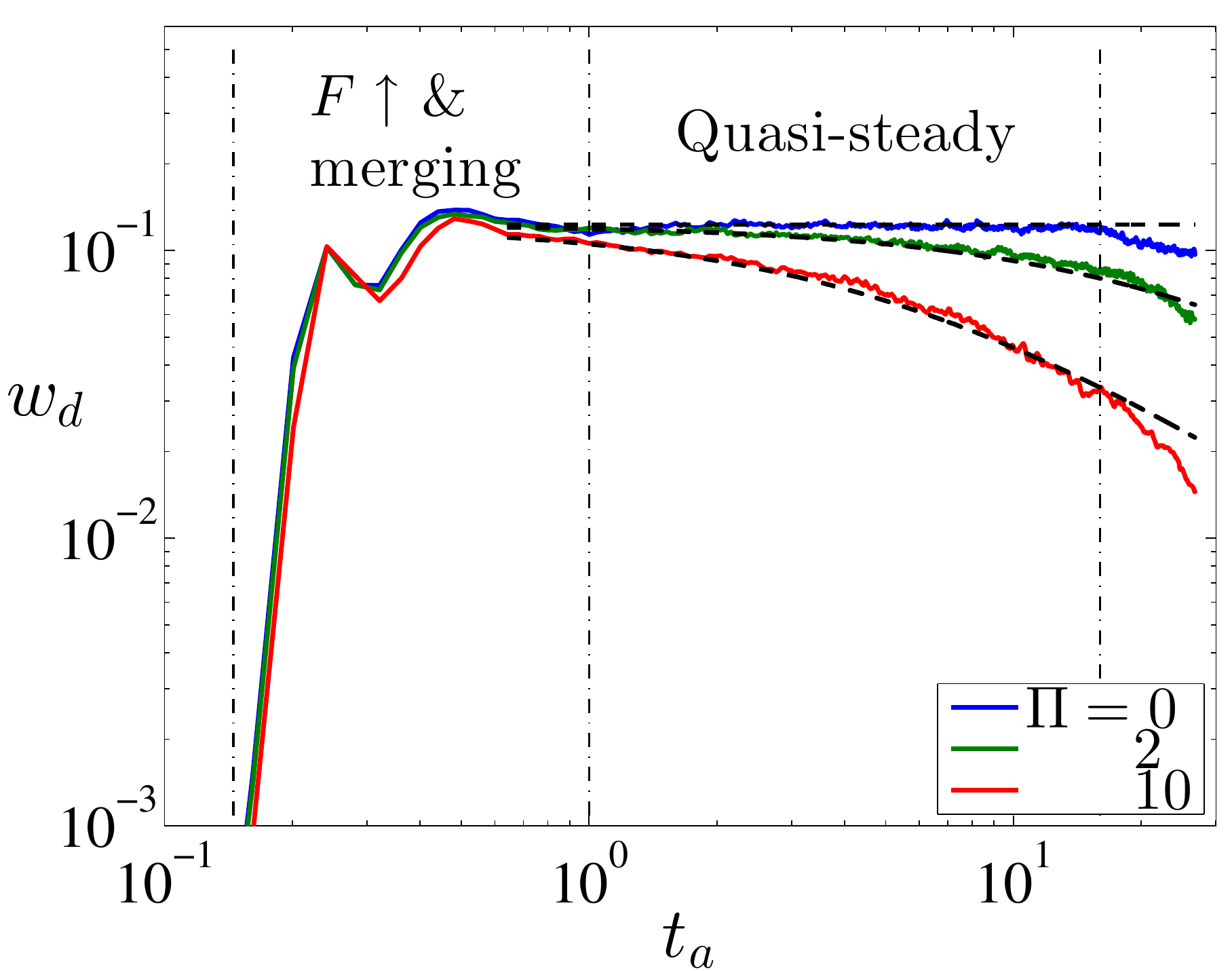}}\par
}
\caption{Numerical results on $z = -0.01$ at $Ra_0 = 20000$ for different $\param$: evolution in time of ($a$) the concentration profile $C$; ($b$) the horizontal-mean finger width $\delta$; and ($c$) the magnitude of horizontal-mean downward velocity $w_d$.  In ($a$), only a small portion of $x$ is shown and the two (white) dashed-dot lines indicate the times of transition to the flux-growth \& plume-merging regime and the quasi-steady convective regime, respectively, for $\param=0$.  In ($b$), the finger width is measured {\color{black}{using $\delta=\pi/[\overline{(\partial C'/\partial x)^2}/\overline{C'^2}]^{1/2}$ where $C' = (C - \overline{C})$ is the fluctuation of the concentration field (see more details in \citealt{Slim2014})}}.  In ($b$) and ($c$), the dynamical regimes are delineated as in figure~\ref{fig:DNS_Cs_Flux_Ra20000}, and $\delta$ and $w_d$ can be fitted {\color{black}{using the functional of the mathematical model developed later in (\ref{Cs_F_steady_model}) for $C_s$ in the quasi-steady convective regime, i.e. $\delta = 0.015(0.006\param t_a + 1)$ (dashed) and $w_d = {1}/{[8.1(0.0168\param t_a +1)]}$ (dashed).}}  Before transition to the quasi-steady convective regime, the dynamics are generally not affected by $\param$. However, in the quasi-steady convective regime, the increase of $\param$ enhances the reduction of the concentration field near the upper wall, resulting in a less vigorous convection in the boundary layer: the plumes become wider; the descending velocity is decreased; and less ribs (which represent the proto-plumes) exist in the `fish-bone' pattern and merge with the primary finger roots.} \label{fig:Results_zp01}
\end{figure}

Convection in an open system exhibits a succession of different flow regimes defined by the behaviour of the solute flux \citep{Riaz2006,Tilton2014,Slim2014}. In this study, we distinguish the following regimes defined in figure~\ref{fig:DNS_Cs_Flux_Ra20000}($a$): an initial `diffusion dominant' regime, followed by the `flux-growth \& plume-merging' and `quasi-steady' convective regimes, and the final `shut down' of convection. Below we consider the effect of dissolution capacity, $\param$, on these regimes in turn and show that the effect increases with time.

The `diffusion dominant' regime is not significantly affected by $\param$, due to the early onset of convection at high $Ra_0$. This prevents the reduction of $C_s$ at the interface (see figure~\ref{fig:DNS_Cs_Flux_Ra20000}{\emph{b}}), so that the flux exhibits a diffusive decay, $F \sim (\pi t)^{-1/2}$. This behaviour continues up to {\color{black}{$t_a = 3000/Ra_0$ (i.e. $t_{ad} \approx 3000$)}} even after perturbations have begun to grow linearly, since the nascent fingers are still encompassed within the relatively thick diffusive boundary layer.
 
During the `flux-growth \& plume-merging' regime the boundary layer scallops and the penetration of unsaturated fluid to the interface increases the flux to a maximum. The evolution of the finger root concentration in figure~\ref{fig:Results_zp01}(\emph{a}) shows that fingers start to travel laterally, which leads to merging of neighbours and a coarsening of the pattern. Due to the short time scales, the basic flow characteristics, e.g. the horizontal-mean finger width $\delta$ and magnitude of horizontal-mean downward velocity, $w_d$, are still not affected by $\param$ (see figure~\ref{fig:Results_zp01}\emph{b} and \emph{c}).  However, the decrease of the interface concentration $C_s$ at large $\param$ becomes more evident and begins to reduce the solute flux (see figure~\ref{fig:DNS_Cs_Flux_Ra20000}).   

\floatsetup[figure]{style=plain,subcapbesideposition=top}  
\begin{figure}[t]
{
  \centering
  \sidesubfloat[]{\includegraphics[width=0.8\textwidth]{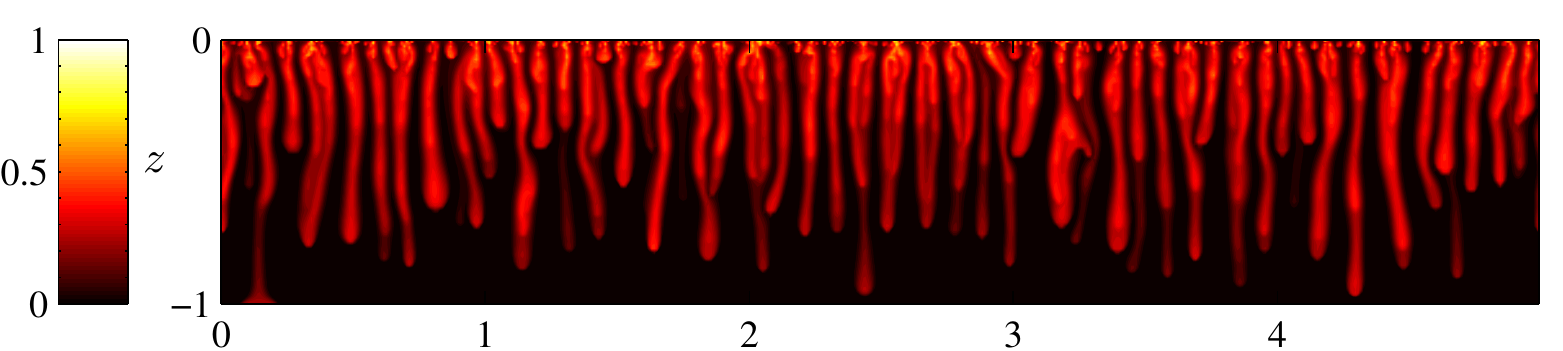}\quad
  			\includegraphics[width=0.127\textwidth]{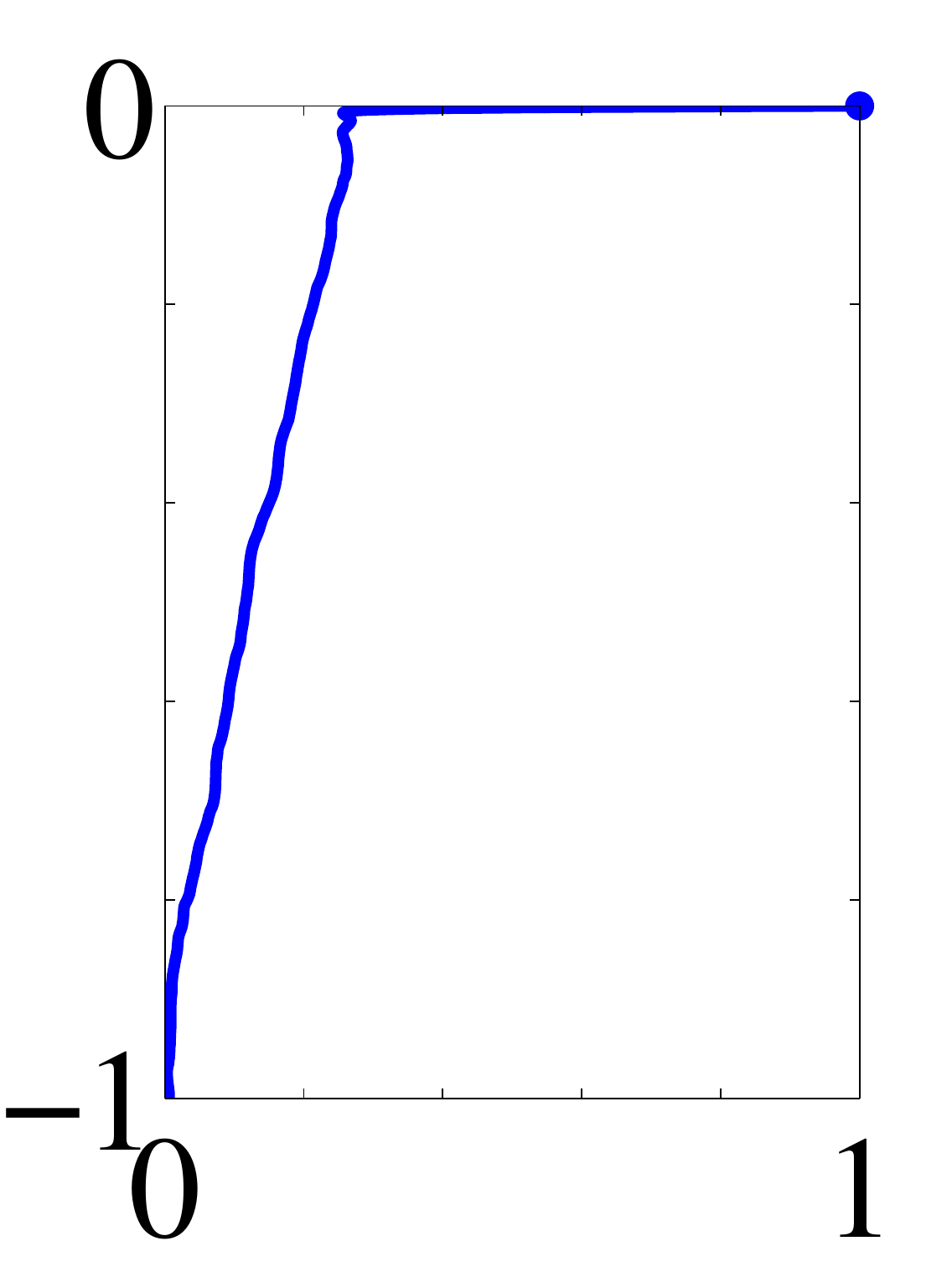}}\\
  \sidesubfloat[]{\includegraphics[width=0.8\textwidth]{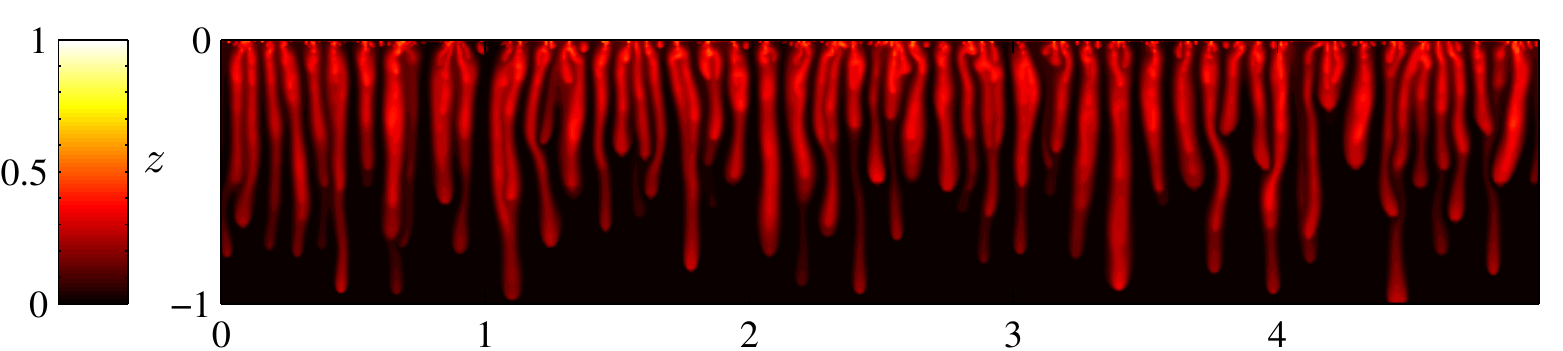}\quad
  			\includegraphics[width=0.124\textwidth]{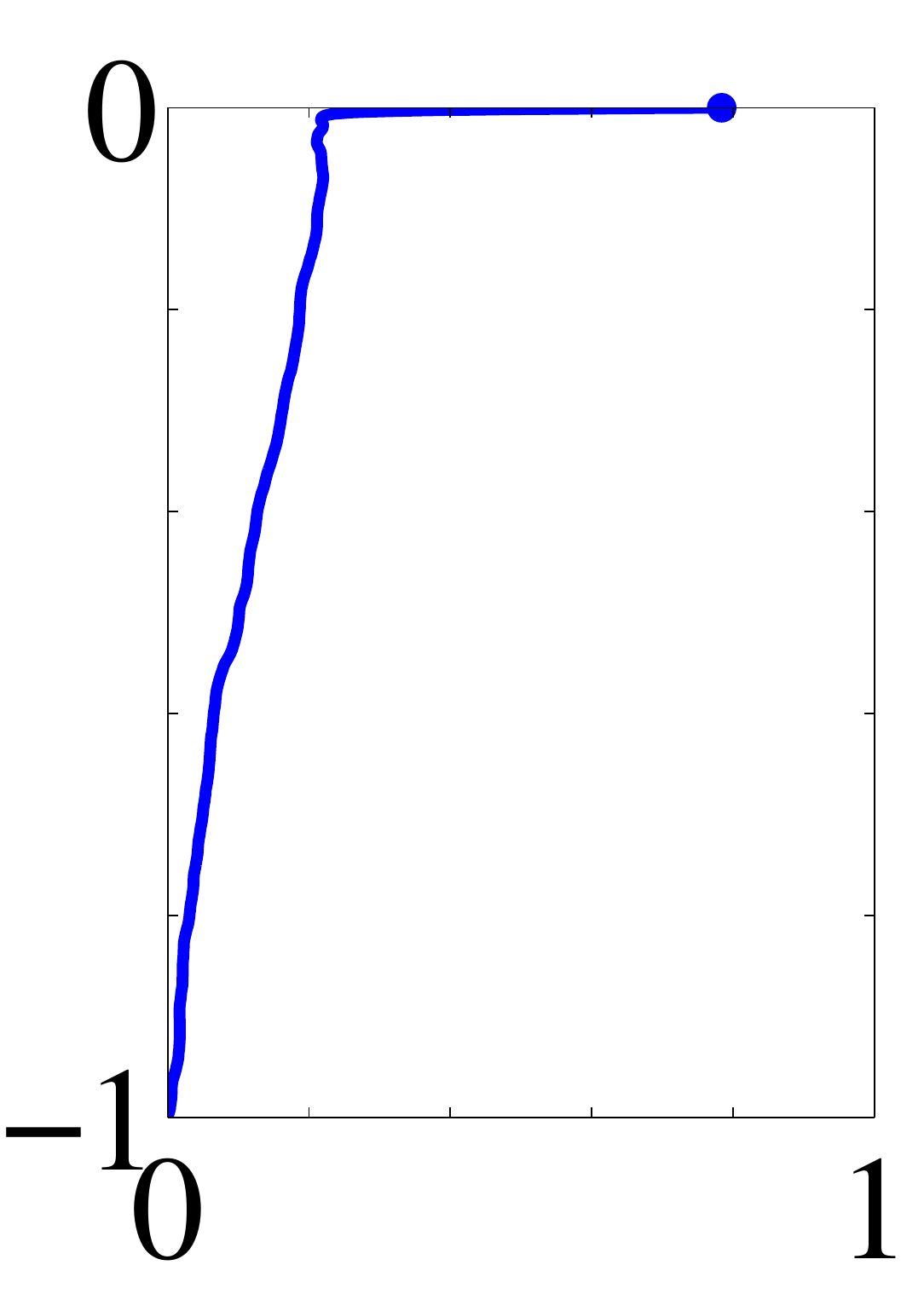}}\\
  \sidesubfloat[]{\includegraphics[width=0.8\textwidth]{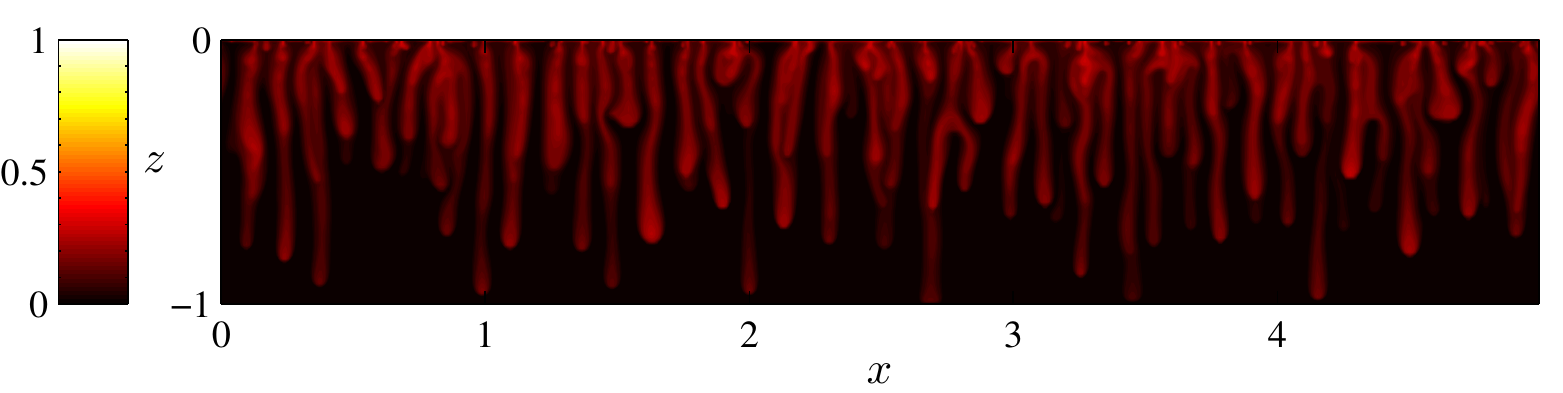}\quad
  			\includegraphics[width=0.126\textwidth]{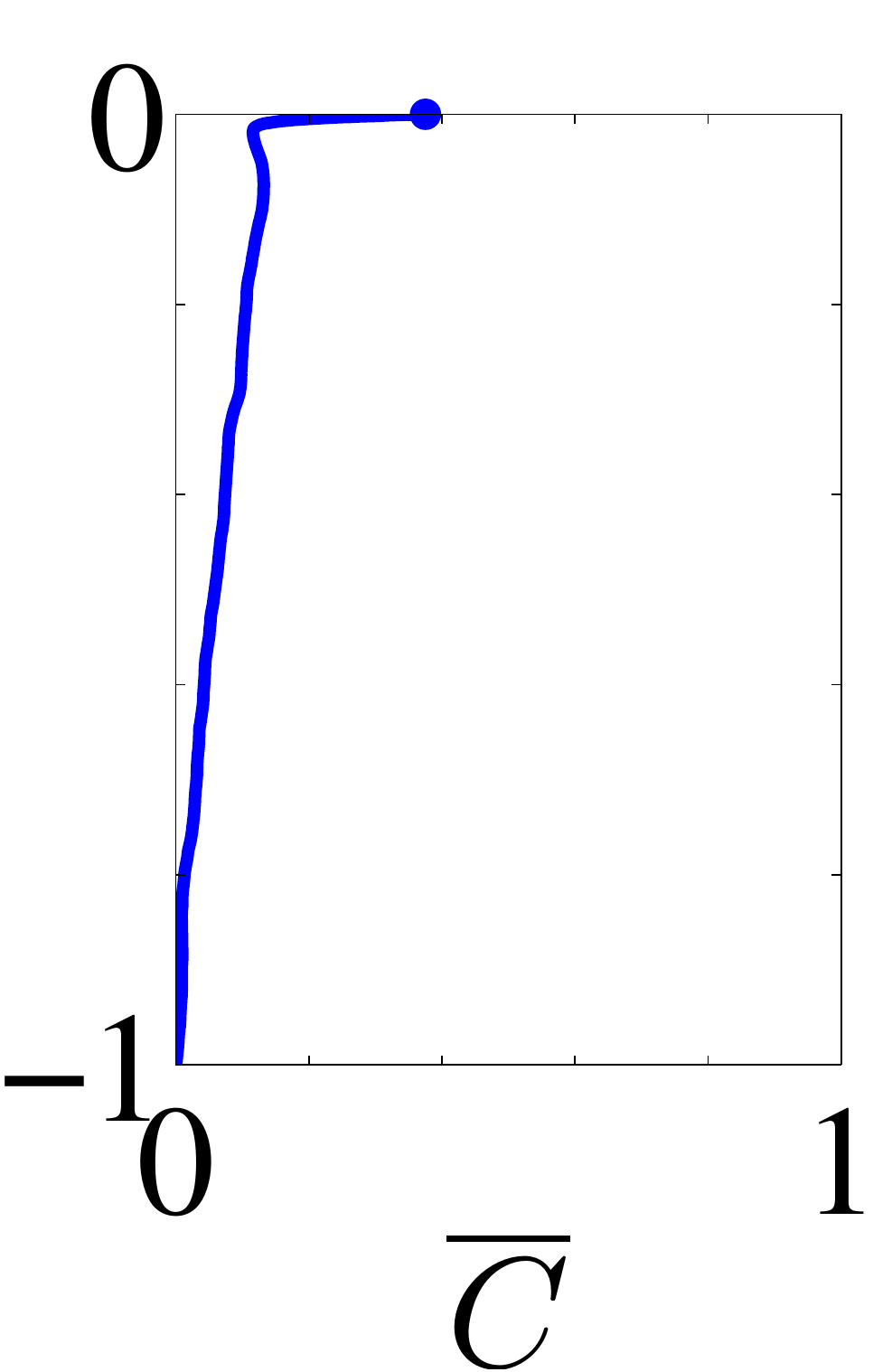}}\par
}
\caption{Snapshots of the concentration field $C$ and the corresponding horizontal-mean concentration profile $\overline{C}$ from DNS at $Ra_0 = 20000$ when the downwelling fingers first reach the bottom wall: ($a$) $\param = 0$, $t_a = 7.2$; ($b$) $\param = 2$, $t_a = 8$; and ($c$) $\param = 10$, $t_a = 10$.  The dot at the top of $\overline{C}$ denotes the concentration at the upper boundary.  The increase of $\param$ reduces the plume-root concentration and decreases the plume descending speed.} \label{fig:Snapshot_tb}
\end{figure}

After the convective pattern has coarsened, the flow transitions to a `quasi-steady' convective regime. At these longer timescales ($t_a \approx 1$), $C_s$ begins to drop rapidly (see figure~\ref{fig:DNS_Cs_Flux_Ra20000}\emph{b}) and the difference between convection in open and closed systems is most pronounced. The dissolution flux in an open system is constant, while the flux in a closed system decays ever more rapidly with increasing $\param$  (see figure~\ref{fig:DNS_Cs_Flux_Ra20000}\emph{a}). Therefore, despite the name of the convective regime, convection in a closed system is never actually quasi-steady. In both open and closed systems, small proto-plumes are continuously generated at the top boundary, swept sideways, and assimilated into the large fingers that penetrate to greater depth. This generates a typical fish-bone pattern in the evolution of the finger root concentration (\citealp{Hewitt2012}, also see figure~\ref{fig:Results_zp01}\emph{a}) and a columnar large-scale flow pattern in the interior (see figure~\ref{fig:Snapshot_tb}). In a closed system, the drop in $C_s$ with time reduces the finger-root concentration and the generation of proto-plumes from the upper wall. This leads to a characteristic `fading fish-bone pattern' for convection in closed systems.  As the driving force for convection declines, the wavelength of the large-scale flow pattern coarsens and the downwelling plumes slow down (see figure~\ref{fig:Results_zp01}$b$ and $c$).

\floatsetup[figure]{style=plain,subcapbesideposition=top}  
\begin{figure}[t]
{
  \centering
  \sidesubfloat[]{\includegraphics[width=0.8\textwidth]{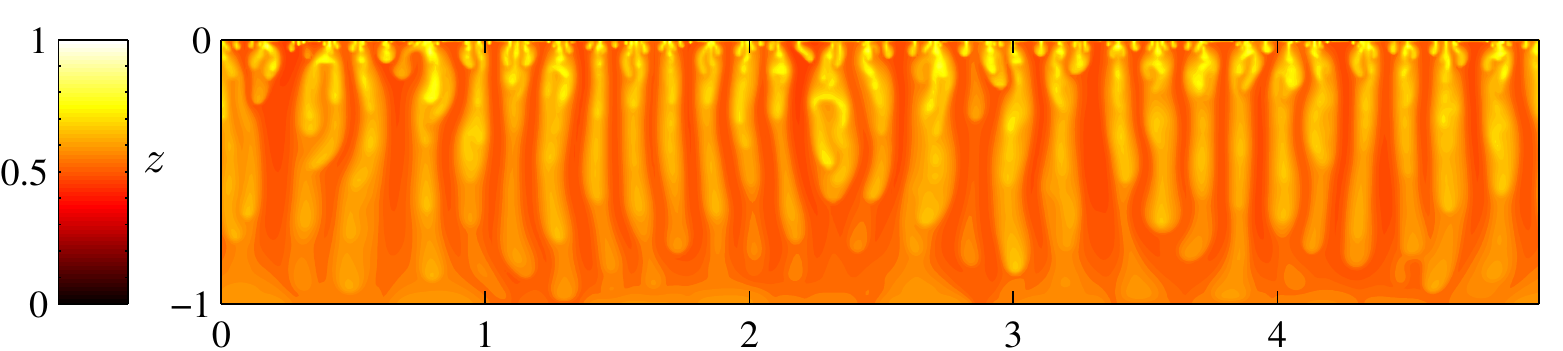}\;\;\hspace{-0.0in}
  			\includegraphics[width=0.1235\textwidth]{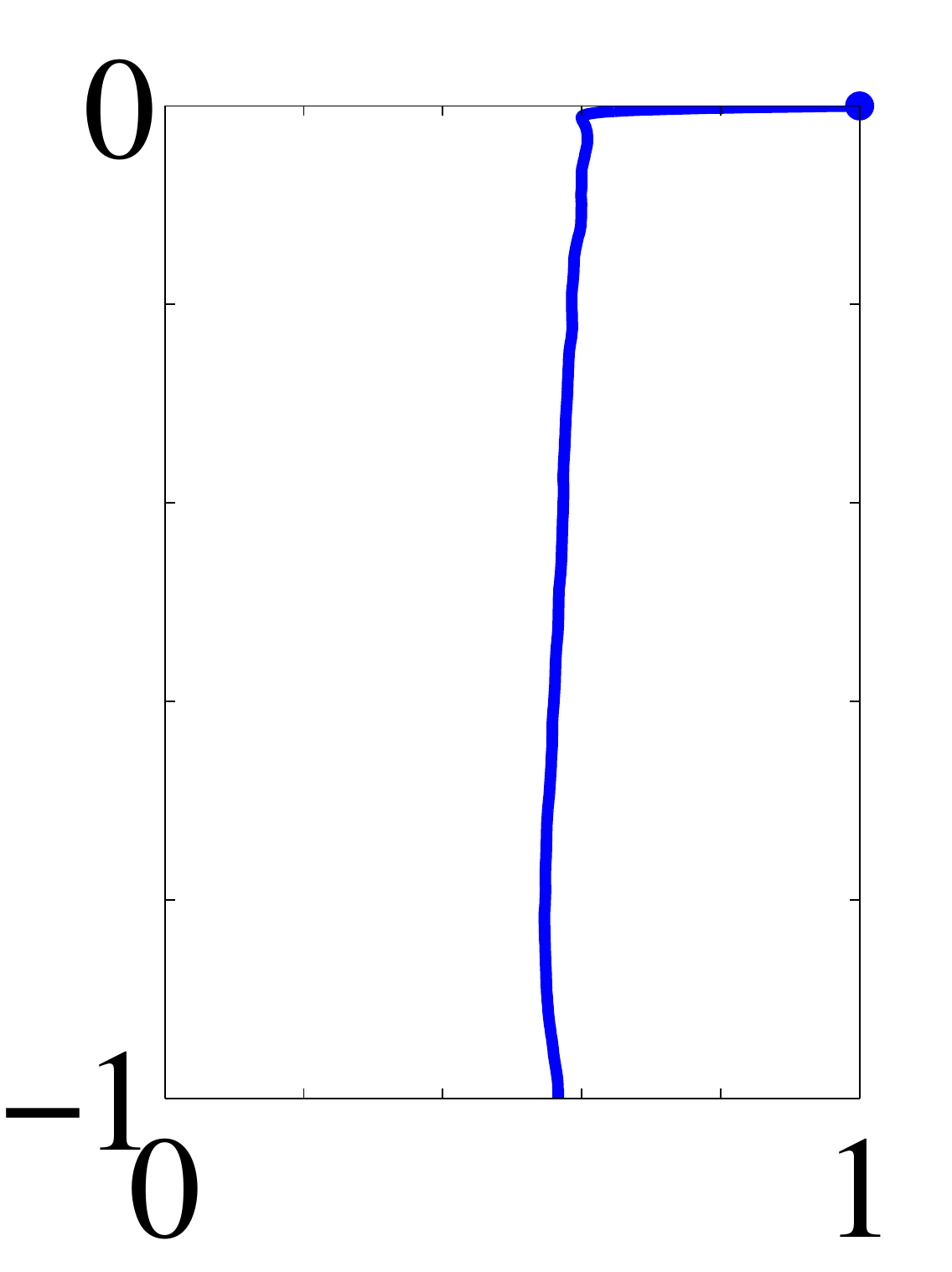}\;}\\
  \sidesubfloat[]{\includegraphics[width=0.8\textwidth]{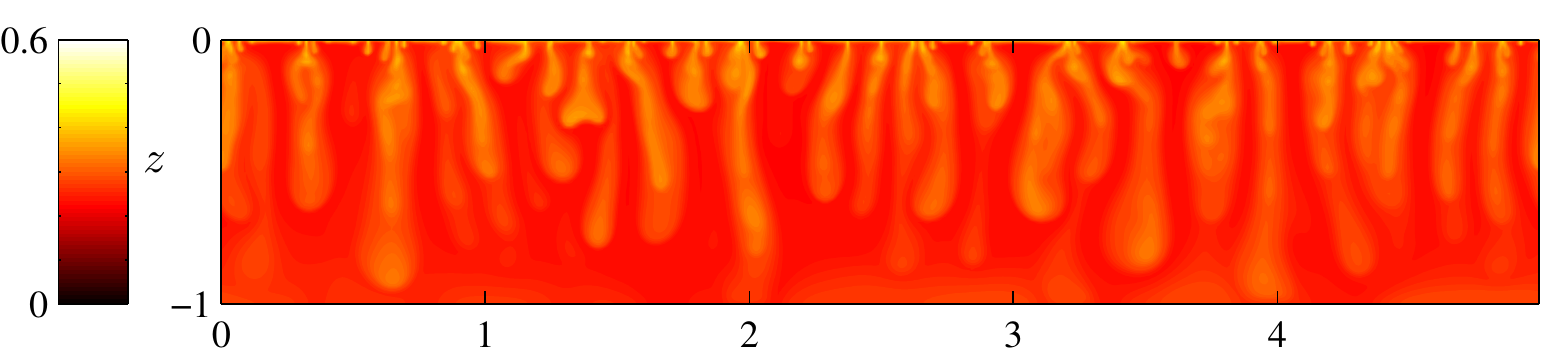}\quad
  			\includegraphics[width=0.131\textwidth]{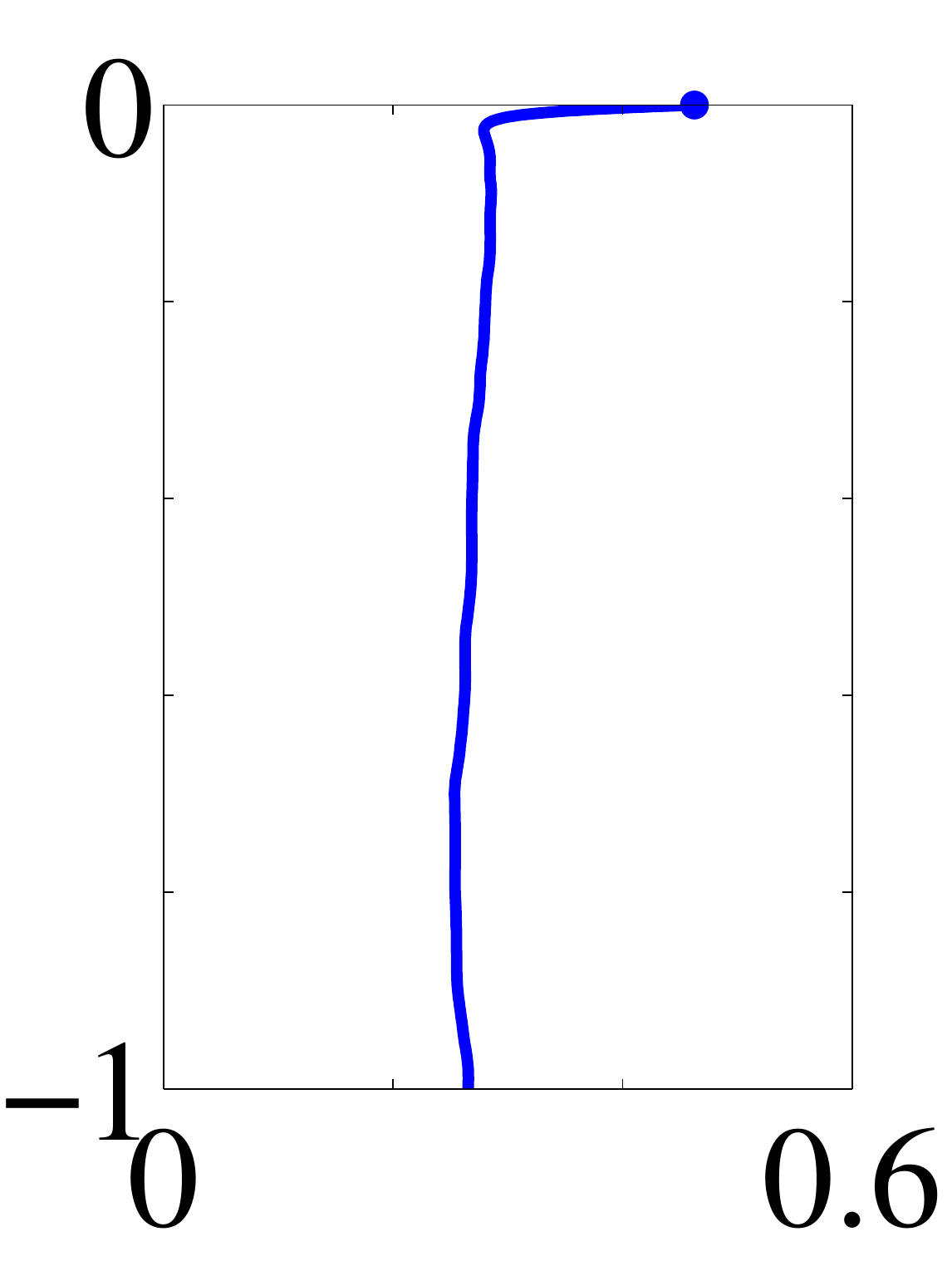}}\\
  \sidesubfloat[]{\includegraphics[width=0.8\textwidth]{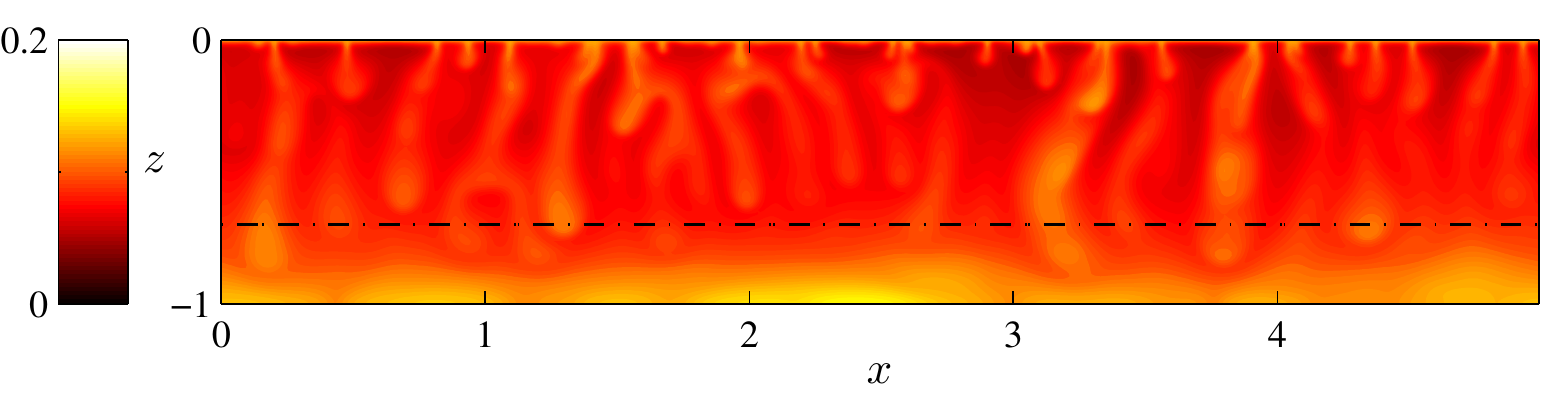}\quad
  			{\includegraphics[width=0.131\textwidth]{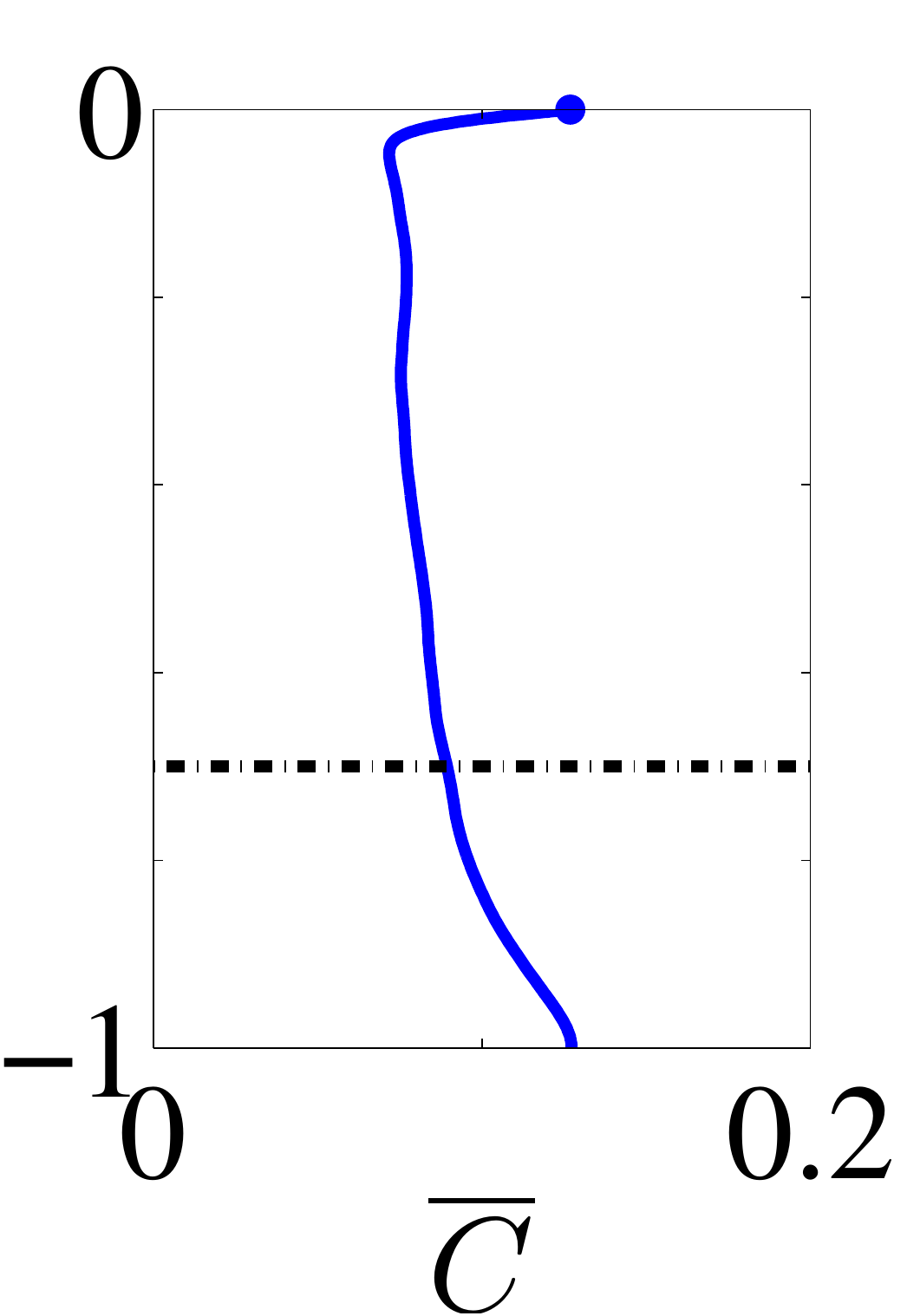}}}\par
}
\caption{Snapshots of the concentration field $C$ and the corresponding horizontal-mean concentration profile $\overline{C}$ from DNS at $Ra_0 = 20000$ and $t_a = 50$: ($a$) $\param = 0$; ($b$) $\param = 2$; and ($c$) $\param = 10$.  The dot at the top of $\overline{C}$ denotes the concentration at the upper boundary.  When convection is shut down, the horizontal-mean concentration profile outside the top diffusive boundary layer becomes nearly independent of $z$.  However, at large $\param$, e.g. $\param = 10$ shown in ($c$), due to the rapid decline of $C_s$ at the interface, this system exhibits a three-layer dynamics: above the dashed-dot line is the regular, two-layer shut-down convection as in ($a$, $b$); beneath the dashed-dot line, the fluid is stably stratified and the flow is mainly by diffusion.} \label{fig:Snapshot_shutdown}
\end{figure}

After the fingers reach the lower boundary, the CO$_2$-rich fluid starts to move upwards with the returning flow. Once this dense fluid reaches the upper boundary, the driving force for convection is decreased, the flux declines rapidly, and eventually the convection is shut down (see figure~\ref{fig:Snapshot_shutdown}).  Previous work on convective shut down, for  $\param = 0$, shows that
the horizontal mean concentration  is well-mixed and almost constant with depth, outside the diffusive boundary layer at the top (\citealt{Hewitt2013shutdown, Slim2013, Slim2014}, also see figure~\ref{fig:Snapshot_shutdown}$a$), namely, 
\begin{eqnarray}
	\overline{C} \approx \overline{\overline{C}}(t). \label{Cbar_shutdown}
\end{eqnarray}
Based on this observation, theoretical box models were developed  to predict the variation of the dissolution flux in time for open systems.  In this study, our simulation results indicate that (\ref{Cbar_shutdown}) is still valid for {\color{black}{$0 < \param \le 5$}}. As shown in figure~\ref{fig:Snapshot_shutdown}, however, for $\param \ge 10$ the mean concentration profile exhibits a three-layer structure due to the rapid decrease of $C_s$: near the upper wall is the thin diffusive boundary layer; in the core $\overline{C}$ is nearly independent of $z$; and near the bottom wall the fluid is stably stratified.  In the following section, we will extend these theoretical box models to closed systems that do not form such a stable stratification at the base. 

\floatsetup[figure]{style=plain,subcapbesideposition=top}  
\begin{figure}[t]
{ \centering{
  \sidesubfloat[]{\includegraphics[height=1.9in]{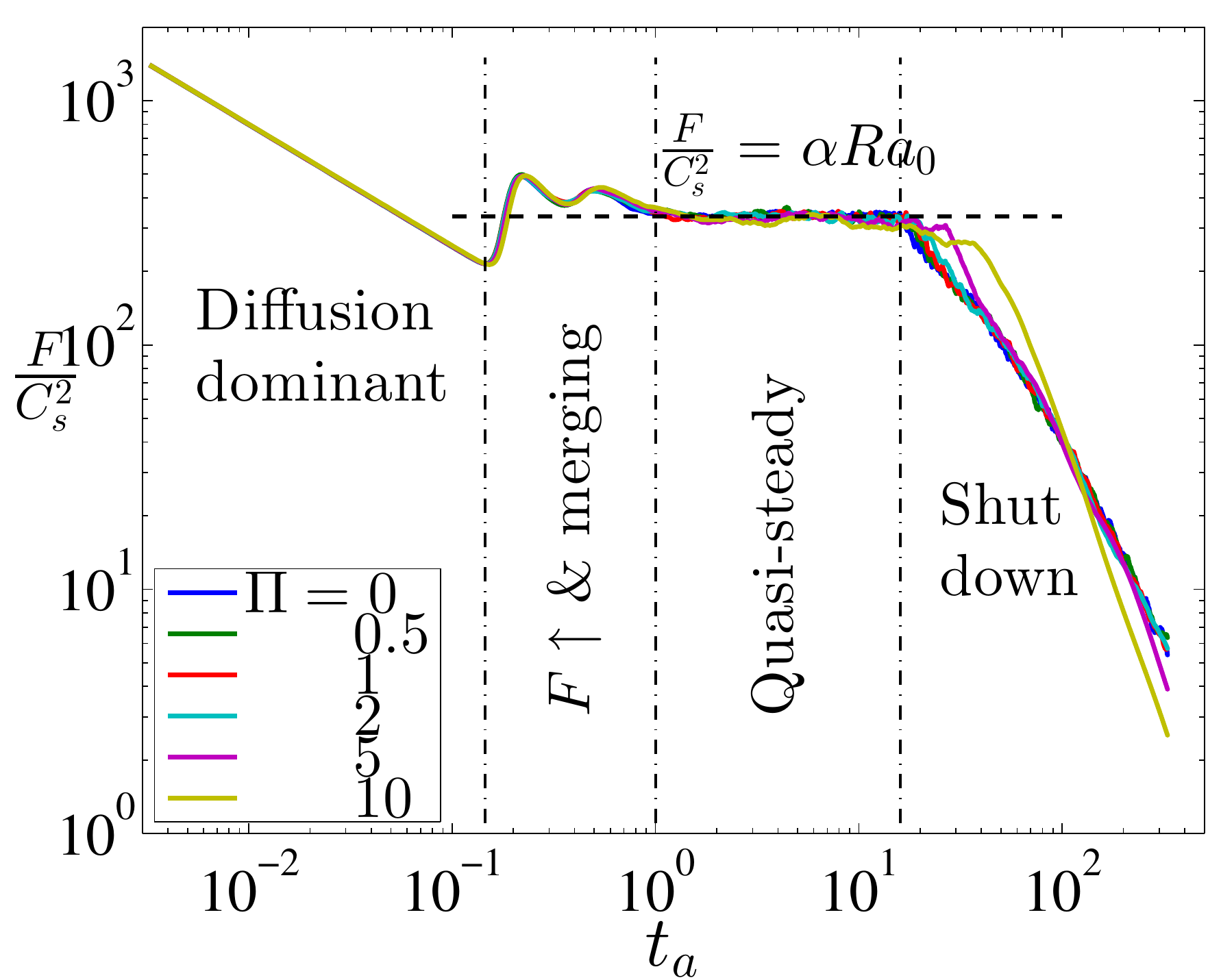}} \quad
  \sidesubfloat[]{\includegraphics[height=1.9in]{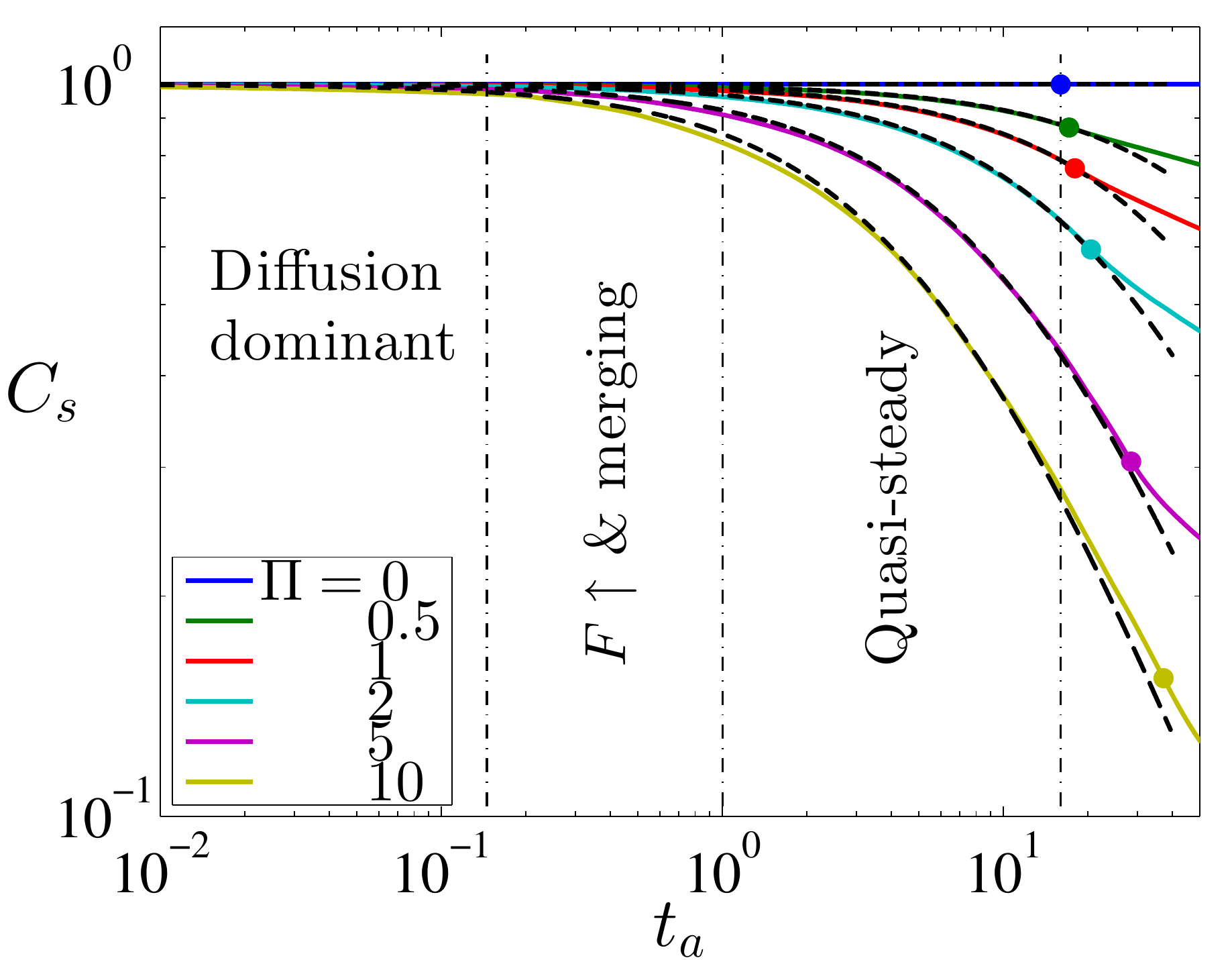}} \\ 
  \sidesubfloat[]{\includegraphics[height=1.9in]{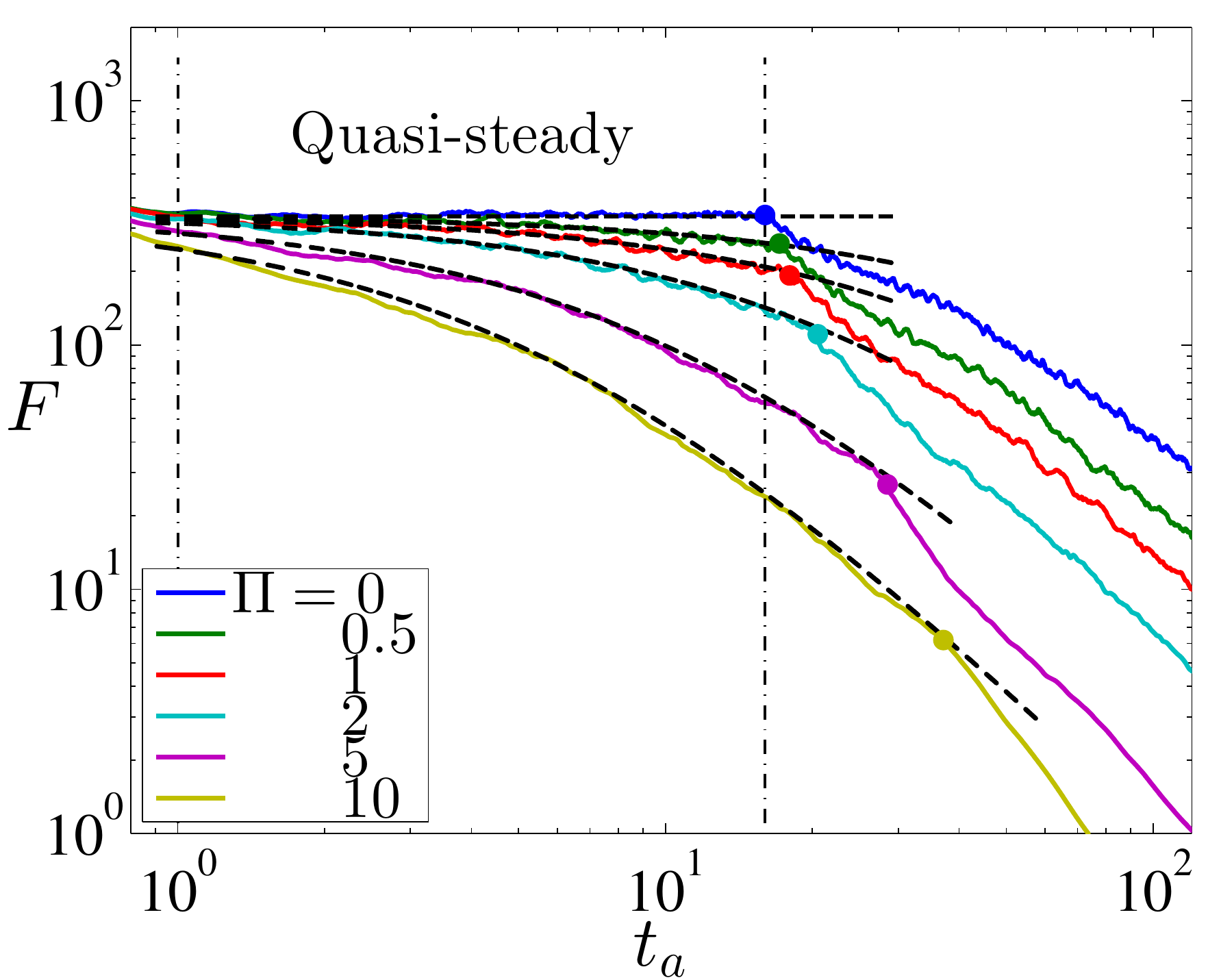}} \quad
  \sidesubfloat[]{\includegraphics[height=1.9in]{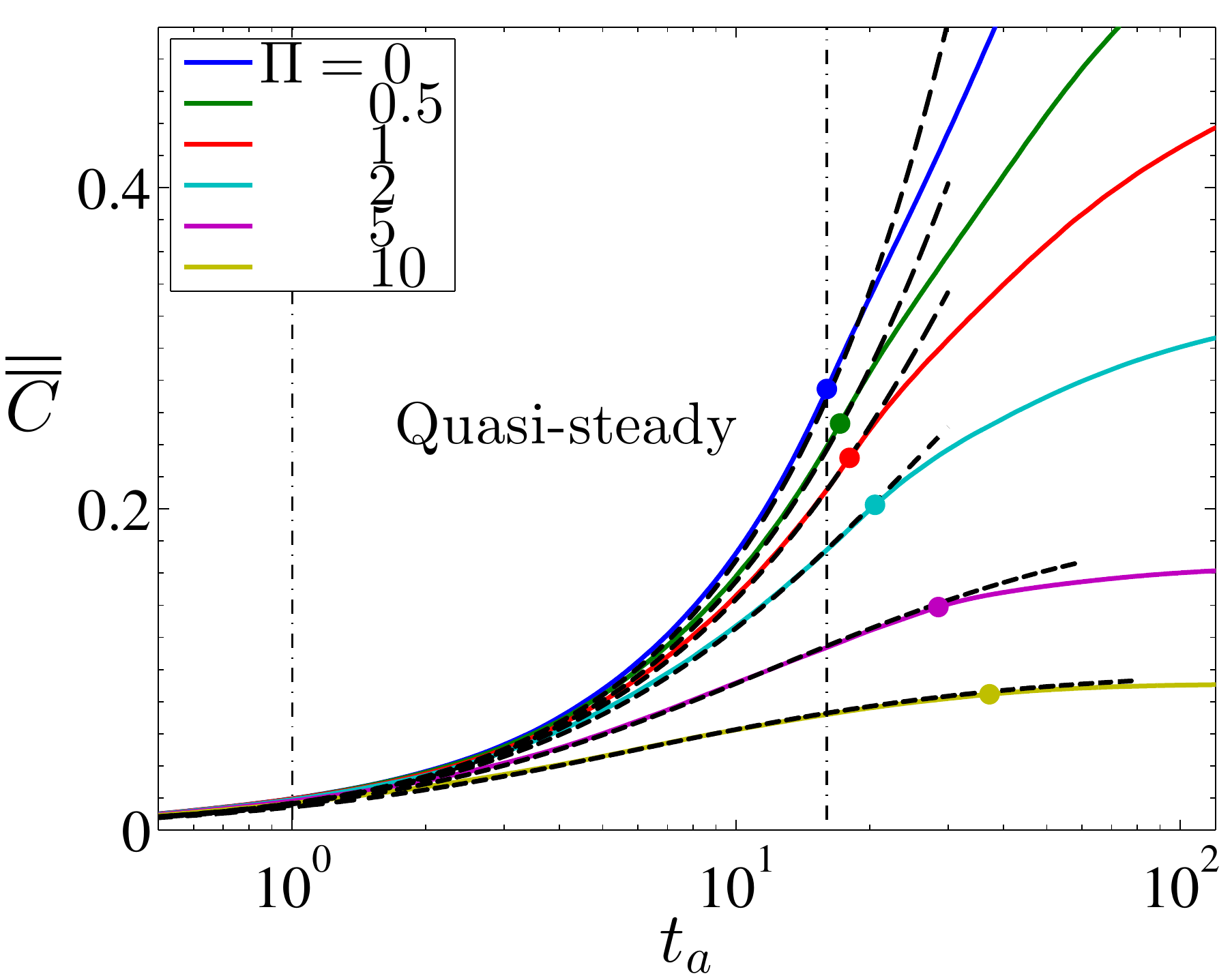}} \par
  }
}
\caption{Mathematical models for the quasi-steady convective regime in closed systems at $Ra_0 = 20000$: ($a$) evolution of the rescaled dissolution flux $F/C_s^2$ in time; ($b$--$d$): comparisons of the interface concentration $C_s$, the dissolution flux $F$ and the volume-averaged concentration $\overline{\overline{C}}$ between the DNS results (solid lines) and the mathematical models (dashed lines).   The dynamical regimes are delineated as in figure~\ref{fig:DNS_Cs_Flux_Ra20000}.  In the quasi-steady convective regime, the rescaled dissolution flux keeps constant in time and is independent of $\param$, i.e. $F/C_s^2 = \coeff Ra_0$ where $\coeff = 0.0168$.  In ($b$)--($d$), the dots mark the time of transition to the shut-down regime {\color{black}{from DNS}} for various $\param$.  Obviously, this transition is delayed with increasing $\param$.} \label{fig:DNSvsModel_quasisteady}
\end{figure}

\subsection{Simple mathematical models}\label{sec:models}
Here we aim to develop a zero-dimensional representations for the convecting system that capture the evolution of the averaged system quantities, e.g. $C_s(t)$, $F(t)$ and $\overline{\overline{C}}(t)$, in different regimes. In the open system this is possible, since the quasi-steady flux in high-$Ra_0$ convection can be expressed as a power law of the form
\begin{eqnarray}
	F = \coeff\, {Ra_0}^{\expo} \quad \mathrm{for} \quad Ra_0 > Ra^*,	 \label{Flux_open}
\end{eqnarray}
where $Ra^* \approx 2000$ gives the onset of the power-law scaling for the one-sided penetrative convection considered here \citep{Slim2014}. Our simulations give the following coefficients, $\coeff = 0.0168$ and $\expo = 1$. Similar values for $\coeff$ and $\expo$ have been found in previous investigations of convection in porous media \citep{Doering1998, Otero2004, Pau2010, Hidalgo2012, Hewitt2012, Elenius2012, Slim2014, Wen2012, Wen2013, Wen2015JFM, WenChini2018JFM}, although some authors have argued for $\expo<1$ \citep{Neufeld2010,Backhaus2011}.

Section~\ref{sec:DNSresults}, however, shows that no such power-law scaling exists for closed systems, as the flux in the quasi-static regime is not constant (see figure~\ref{fig:DNS_Cs_Flux_Ra20000}\emph{a}). Nevertheless, from (\ref{Solute_nondim}) the downward flux beneath the upper diffusive boundary layer at high $Ra_0$ is largely advective and given by 
\begin{eqnarray}
	F \approx -Ra_0\,\overline{wC}\approx Ra_0 w_d \overline{C}\sim Ra_0\, \overline{C}^2, 	 \label{Flux_advective}
\end{eqnarray}
since the magnitude of the horizontal-mean downward velocity  $w_d \sim C$, as shown by (\ref{Darcy_nondim}). In an open system, the interface concentration, $C_s$, is constant and during the quasi-steady regime, {\color{black}{$\overline{wC}=-\coeff$ from (\ref{Flux_open}) and (\ref{Flux_advective}).}} Although $F$ and $C_s$ vary with time in a closed system, (\ref{Flux_advective}) suggests that $F/C_s^2 \sim Ra_0 $ in the quasi-static regime. Figure~\ref{fig:DNSvsModel_quasisteady}($a$) shows that indeed 
\begin{eqnarray}
	\dfrac{F}{C_s^2} = F\vert_{\param = 0} = \coeff\,Ra_0,	 \label{Flux_rescale_constant}
\end{eqnarray}
for different $\param$, which allows the extension of previous box models to closed systems. Combining (\ref{eq:ODE_non_dim}), (\ref{Flux_nondim}) and (\ref{Flux_rescale_constant}) yields the mathematical models for $F$ and $C_s$ in the quasi-steady convective regime:
\begin{eqnarray}
	C_s(t_a) = \dfrac{1}{\coeff\param t_a + 1} \quad \text{and} \quad F(t_a) = \dfrac{\coeff Ra_0}{\left(\coeff\param t_a + 1\right)^2}. \label{Cs_F_steady_model}
\end{eqnarray}
Moreover, from (\ref{Cs_C_volavg}) the total amount dissolved is given by
\begin{eqnarray}
	\overline{\overline{C}}(t_a) = \dfrac{\coeff t_a}{\coeff\param t_a + 1}. \label{Cint_steady_model}
\end{eqnarray}

As shown in figure~\ref{fig:DNSvsModel_quasisteady}($b$--$d$), comparisons of the mathematical models in (\ref{Cs_F_steady_model}) and (\ref{Cint_steady_model}) and the DNS results show good agreements in the quasi-steady convective regime, for $\param \leq 10$.   In addition, the assumption that $w_d \sim C$ can be confirmed by fitting the data in figure~\ref{fig:Results_zp01}($c$) with an expression of the form (\ref{Cs_F_steady_model}), to show that  $w_d = C_s/8.1$. Similarly, it can be shown that $\delta \sim 1/C_s$ in figure~\ref{fig:Results_zp01}($b$). It should be noted that these are quantities measured near the interface and appropriate prefactors vary with distance from the interface.


In an open system the fingertip sinks with a nearly constant speed in the quasi-steady convective regime \citep{Riaz2006, Hewitt2013shutdown, Slim2014}.   However, for closed systems, $\param>0$, the downward propagation velocity, $w_d$, slows down as the interface concentration, $C_s$, declines. This delays the transitions from the quasi-steady to the shut-down regime, as shown in figures~\ref{fig:Results_zp01}($c$) and \ref{fig:DNSvsModel_quasisteady}($c$). However, since $w_d\sim C_s$ and the decline of $C_s(t)$ is determined by (\ref{Cs_F_steady_model}), so that the fingertip position of the descending plumes is given by
\begin{eqnarray}
	z_{tip}(t_a) = -\int_0^{t_a} w_d\,d\tilde{t}_c = -\dfrac{\ln{(1+\coeff \param t_a)}}{7.2\, \coeff \param}. \label{ztip_model}
\end{eqnarray}
Here  $w_d=C_s/7.2$, as in our simulations the fingers reach the lower boundary at $t_a \approx 7.2$ for $\param=0$. {\color{black}{According to the model, in closed systems the fingers first hit the base of the domain at}}
\begin{eqnarray}
	t_b = \dfrac{e^{7.2\, \coeff\param} - 1}{\coeff\param}, \label{tb_model}
\end{eqnarray}
when $z_{tip}(t_b) = - 1$.
After reaching the base of the domain dense fluid is carried upward by the return flow and once the saturated fluid reaches the interface, convection shuts down rapidly. Due to the symmetry of the downwelling and upwelling regions (see figure~\ref{fig:Snapshot_tb}), mass balance requires that the magnitude of horizontal-mean upwelling velocity is equal to the magnitude of horizontal-mean downwelling velocity at any time, $w_u \approx w_d$. Therefore, one might expect the time required for the transition to shut down, $t_s$, to be given by solving $z_{tip}(t_b) = - 2$. However, even for $\param=0$ this simple estimate is not accurate and we prefer the expression
\begin{eqnarray}
	t_s = \dfrac{e^{16\, \coeffup\, \param} - 1}{\coeffup\,\param}, \label{ts_model}
\end{eqnarray}
where $w_d = C_s/8$ has been used and $\coeffup = 0.8\,\coeff$. These corrections account for delays due to accumulation of dense fluid at the base and for an apparent reduction of the efficiency of the return flux relative to (\ref{Flux_open}) and (\ref{Flux_rescale_constant}).  As shown in figure~\ref{fig:tbs_Ztip}, the estimates for the timescales given by (\ref{tb_model}) and (\ref{ts_model}) agree very well with the DNS results, as long as {\color{black}{$\param \le 5$}}.  At large $\param$, e.g. $\param = 10$, however, the theoretical predictions of $t_b$ and $t_s$ underestimate the timescales determined from the simulations.  This is due to the formation of a stable density stratification at the base of the domain, shown in figure~\ref{fig:Snapshot_shutdown}(\emph{c}).


\floatsetup[figure]{style=plain,subcapbesideposition=top}  
\begin{figure}[t]
{
  \centering
  \sidesubfloat[]{\includegraphics[height=1.9in]{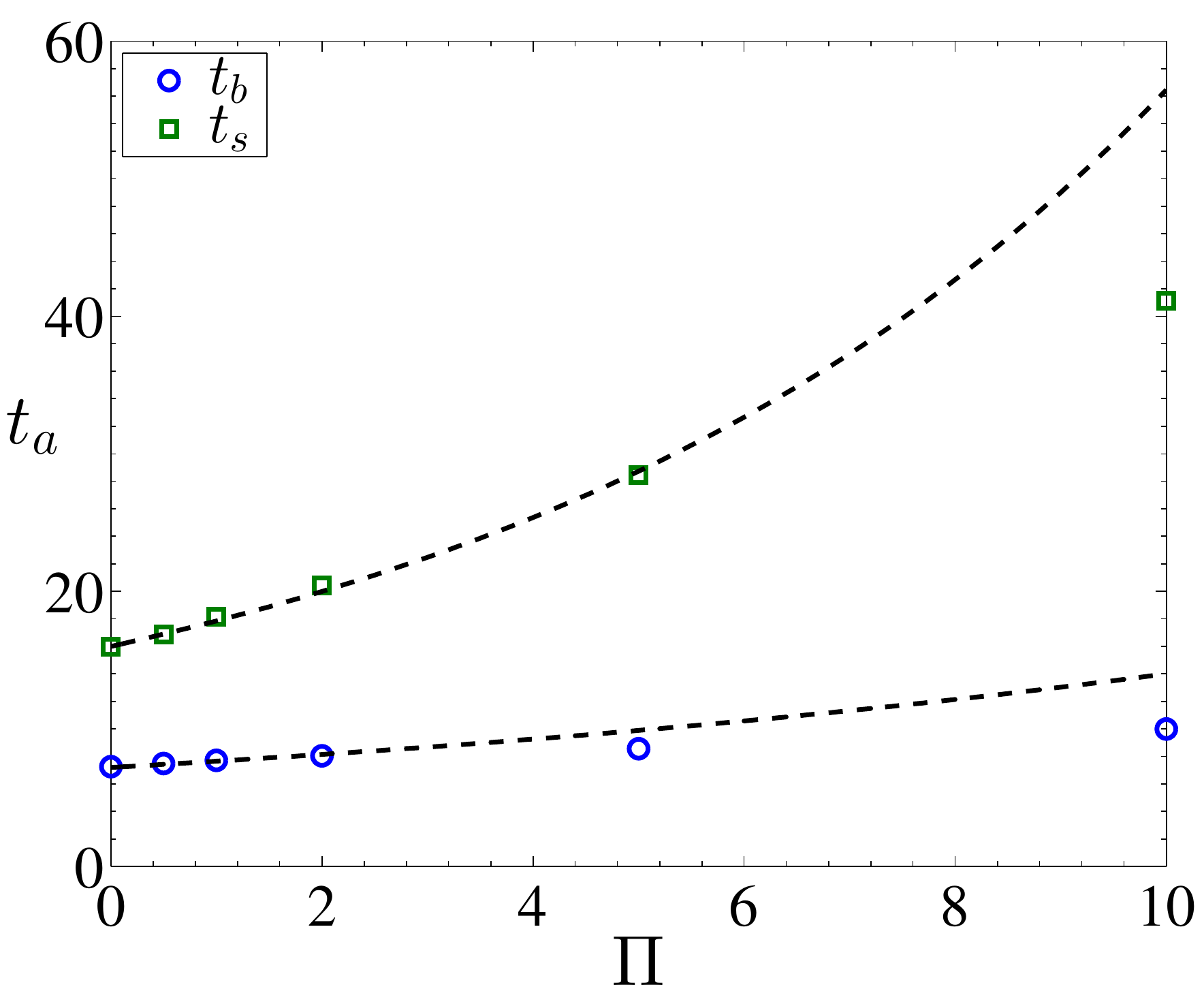}} \quad
  \sidesubfloat[]{\includegraphics[height=1.9in]{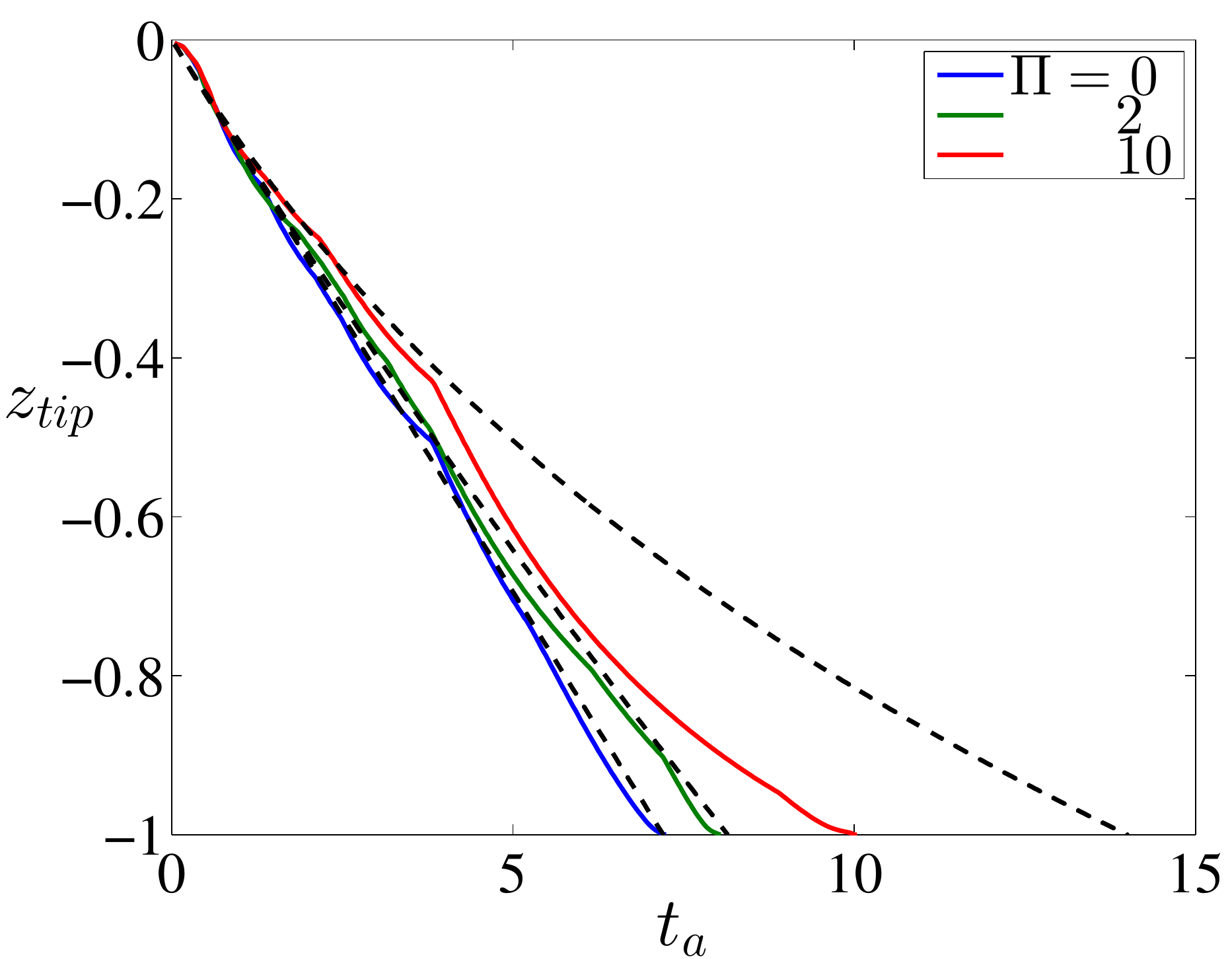}} \par
}
\caption{Measures of finger motions at $Ra_0=20000$ for different $\param$: ($a$) variations of $t_b$ and $t_s$ with $\param$; ($b$) evolution of the fingertip location $z_{tip}$ in time.  Symbols and solid lines: DNS results; dashed lines: mathematical models. In ($b$),  the fingertip location is defined as minima of the $C = 0.05$ contour.}
\label{fig:tbs_Ztip}
\end{figure}


For simulations with {\color{black}{$\param \leq 5$}}, the horizontal mean concentration $\overline{C}$ exhibits a vertically well-mixed structure in the shut-down regime (see figure~\ref{fig:Snapshot_shutdown}).  Therefore, from the definitions in (\ref{Flux_nondim}) and (\ref{C_volavg}) and the approximation in (\ref{Cbar_shutdown}), the dissolution flux of CO$_2$ can be rewritten as 
\begin{eqnarray}
	F = \dfrac{d}{dt} \Large{\int}_{-1}^0\overline{C}dz = \dfrac{d\overline{\overline{C}}}{dt}. \label{Flux_nondim2}
\end{eqnarray}
As in \cite{Hewitt2013shutdown}, we define a \emph{time-dependent} Nusselt number by scaling the flux $F(t)$ up to a unit concentration difference:
\begin{eqnarray}
	Nu(t) = \dfrac{F}{C_s - \overline{\overline{C}}}, \label{Nu}
\end{eqnarray}
where $Nu$ varies as a function of current Rayleigh number, i.e. $Nu(t) = \mathcal{N}(Ra(t))$.  Note that in closed systems $C_s$ also varies as a function of time.  Analogous to high-$Ra$ Rayleigh--B\'{e}nard convection in porous media where the Nusselt number $Nu_{RB}$ linearly depends on a relative Rayleigh number, the Nusselt number $Nu(t)$ in the solutal convection problem can be expressed as 
\begin{eqnarray}
	Nu(Ra(t)) = Nu_{RB}(Ra_e) = \gamma Ra_e, \label{Nu2}
\end{eqnarray}
where the effective Rayleigh number
\begin{eqnarray}
	Ra_e = \chi(C_s - \overline{\overline{C}})Ra_0, \label{Ra_e}
\end{eqnarray}
and $\gamma$ and $\chi$ are two constant numbers.  In convection-time framework, combining (\ref{Flux_nondim2})--(\ref{Ra_e}) with (\ref{Cs_C_volavg}) results in 
\begin{eqnarray}
	\dfrac{d\overline{\overline{C}}}{dt_a} = \gamma \chi \left[1 - (1+\param)\overline{\overline{C}}\right]^2. \label{ODE_shutdown}
\end{eqnarray}
Solving this ordinary differential equation gives 
\begin{eqnarray}
	\overline{\overline{C}}(t_a) = \dfrac{1}{1+\param}\left[1 - \dfrac{1}{\gamma \chi(1+\param) t_a + c_0}\right]. \label{Cint_shutdown_model}
\end{eqnarray}
From (\ref{Cs_C_volavg}) and (\ref{Flux_nondim2}), we obtain the models for $C_s$ and $F$ for the shut-down regime:
\begin{eqnarray}
	C_s(t_a) = \dfrac{1}{1+\param}\left[1 + \dfrac{\param}{\gamma \chi(1+\param) t_a + c_0}\right] \;\; \text{and} \;\; F(t_a) = \dfrac{\gamma \chi Ra_0}{\left[\gamma \chi(1+\param) t_a + c_0\right]^2}. \label{Cs_F_shutdown_model}
\end{eqnarray}
{\color{black}{We choose $\gamma \chi = 0.0317$ and  $c_0 = 0.861$ by fitting (\ref{Cs_F_shutdown_model}) with the DNS data so that at $\param=0$, our model is consistent with the theoretical box model given by \cite{Slim2014} and $F(t_a=16) = \alpha Ra_0$.}}

Figure~\ref{fig:DNSvsModel_shut} shows the comparisons between the mathematical models and the numerical simulations in the shut-down regime for different $\param$ at $Ra_0 = 20000$.  For $\param < 10$, the models in (\ref{Cint_shutdown_model}) and (\ref{Cs_F_shutdown_model}) are in good agreement with the DNS results in the shut-down regime.  {\color{black}{For $\param \ge 10$, however, the model breaks down in the shut-down regime, because a stable stratification forms at the base of the domain. Nevertheless, in these cases the water is already 95\% saturated, so that the additional dissolution during the shut-down regime is negligible. In these cases, the drop in $C_s$ and hence in $Ra_{e}$ is so rapid that convection is not vigorous enough to maintain a well-mixed solution near the bottom.}}

\begin{figure}[t]
{
  \centering
  \includegraphics[width=0.8\textwidth]{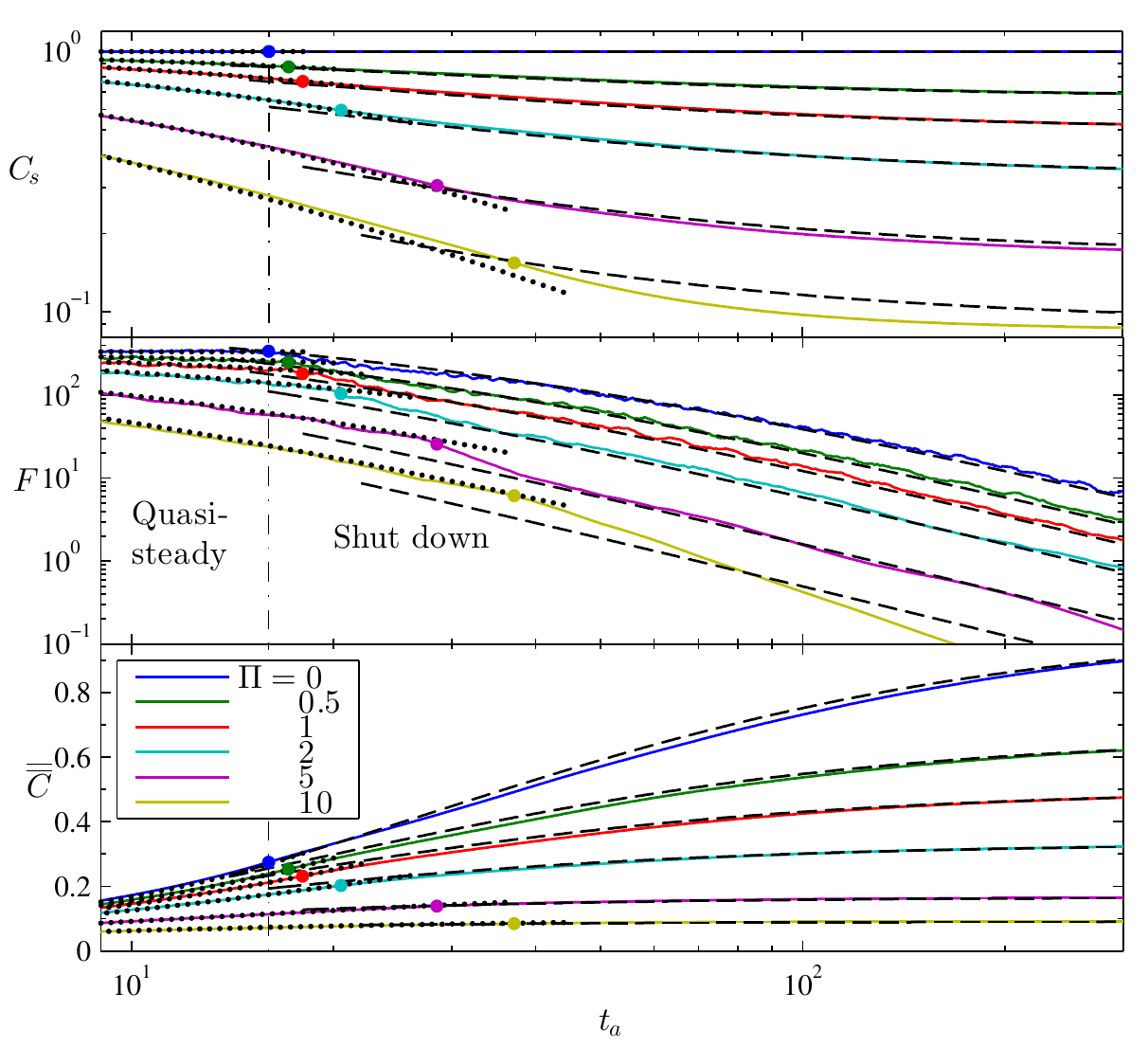}
}
\caption{Comparisons between mathematical models (dashed lines) and numerical simulations (solid lines) for the shut-down regime in closed systems at $Ra_0 = 20000$: ($a$) the interface concentration $C_s$; ($b$) the dissolution flux $F$; and ($c$) the volume-averaged concentration $\overline{\overline{C}}$.  The dots mark the time of transition to the shut-down regime {\color{black}{from DNS}} for various $\param$.  For reference, mathematical models for the quasi-steady regime (dot lines) are also plotted.} \label{fig:DNSvsModel_shut}
\end{figure}
\floatsetup[figure]{style=plain,subcapbesideposition=top}  
\begin{figure}[h!]
{
  \centering
  \sidesubfloat[]{\includegraphics[height=1.9in]{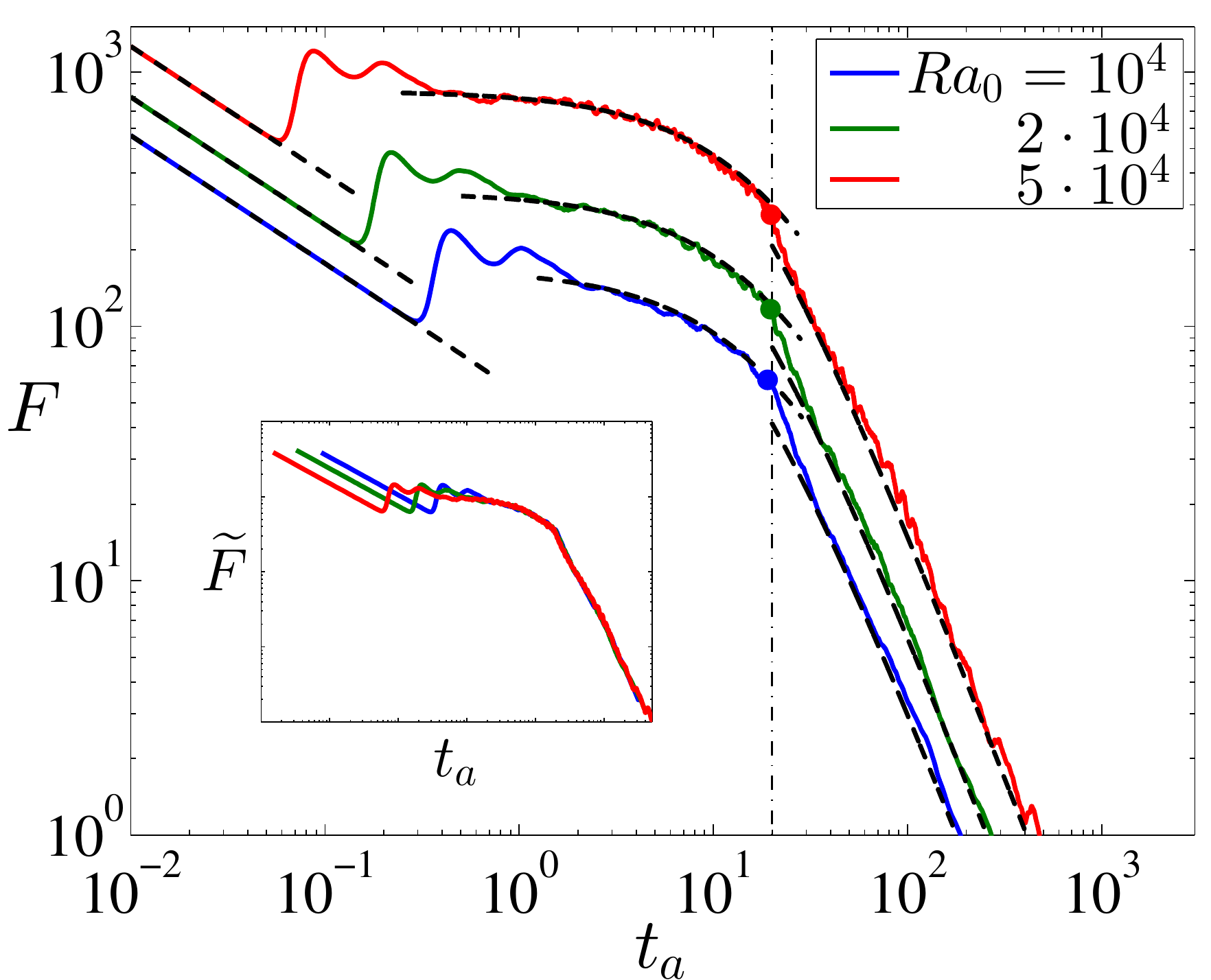}} \quad
  \sidesubfloat[]{\includegraphics[height=1.9in]{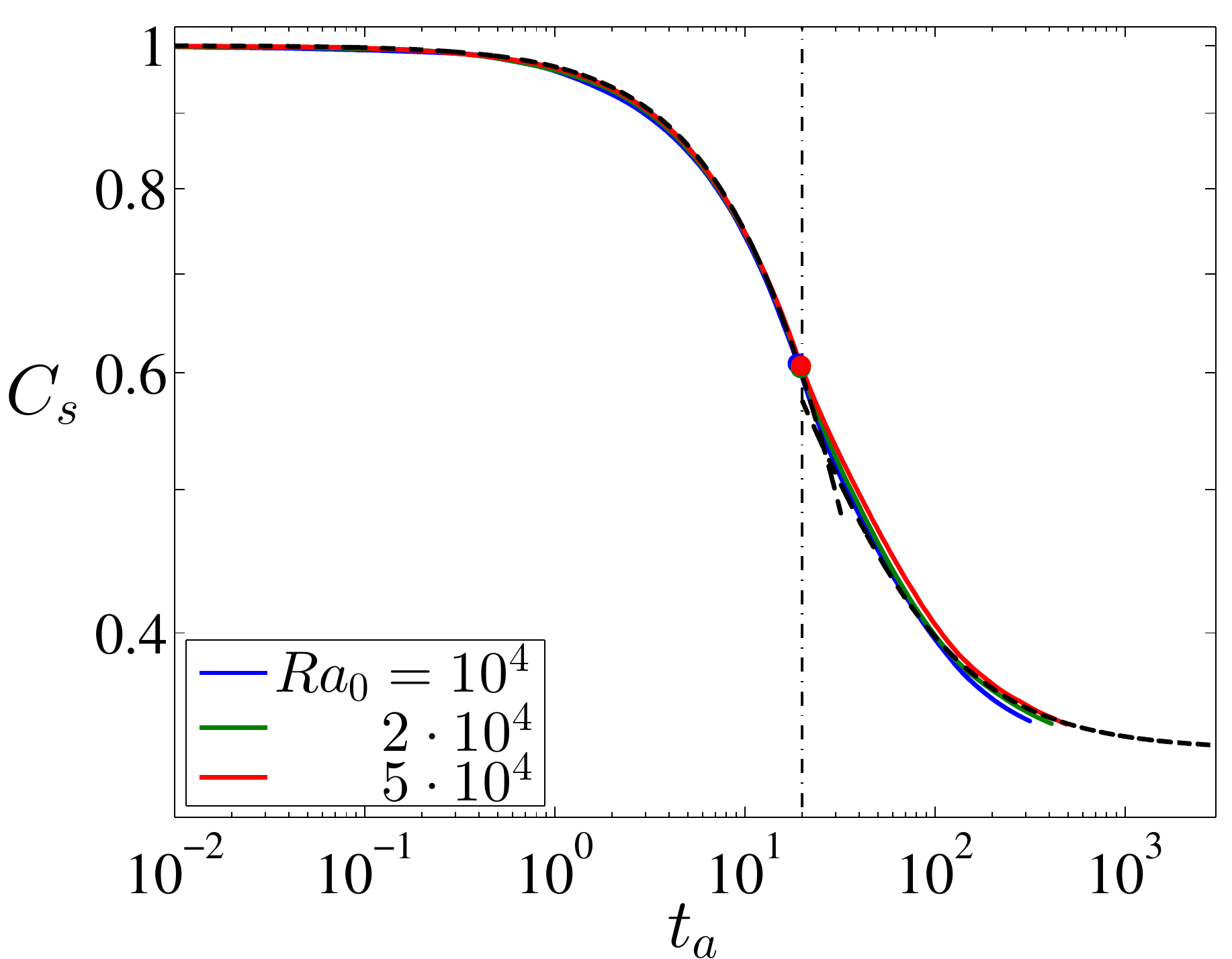}} \par
}
\caption{{\color{black}{Comparisons between mathematical models (dashed lines) and numerical simulations (solid lines)  at $\param = 2$ for various $Ra_0$: ($a$) the dissolution flux $F$; and ($b$) the interface concentration $C_s$.  The inset in ($a$) shows the evolution of normalised flux $\widetilde{F} = F/(\alpha Ra_0)$ in time; and the dots and dashed-dot lines mark the time of transition to the shut-down regime from DNS and the model (\ref{ts_model}), respectively.}}}
\label{fig:DNSvsmodel_eta2}
\end{figure}

{\color{black}{
To verify the mathematical models developed above, DNS were also performed for other high Rayleigh numbers following the same strategy described in the beginning of \S~\ref{sec:DNS_model}, but using different numbers of vertical modes and time steps.  Figure~\ref{fig:DNSvsmodel_eta2} compares the mathematical models with the DNS results in different flow regimes at $\param = 2$ for $Ra_0 = 10000$, 20000 and 50000.  The models (\ref{Cs_F_steady_model}), (\ref{ts_model}) and (\ref{Cs_F_shutdown_model}) match well with the simulation results for various $Ra_0$ due to the asymptotic high-Rayleigh-number behaviour of convection in porous media.  Moreover, for fixed $\param$ our DNS results indeed show that the rescaled dissolution flux $\widetilde{F}$, the time of transition to the shut-down regime $t_s$, and the interface concentration $C_s$ are independent of $Ra_0$ in terms of advection time in both quasi-steady and shut-down regimes, as also revealed from the models. 


Although our models in this manuscript only focus on 2D domains, the study by \cite{Shi2017} reveals that a similar 2D convective modelling strategy predicts the dissolution rate of supercritical \CO2 in a  3D cylinder filled with water-saturated porous media.  Moreover, investigations by \cite{Pau2010}, \cite{Fu2013} and \cite{Hewitt2014} indicate that the power-law-scaling characteristics appearing in 2D also exist in 3D buoyancy-driven porous media convection.  Therefore, it is possible to apply our 2D mathematical models directly to 3D or extend the 2D models to 3D by changing appropriate coefficients.}}

\section{Discussion}\label{sec:discussion}

The pressure drop induced by \CO2 dissolution in a closed reservoir provides a strong negative feedback for convective dissolution. While engineered storage sites are likely open systems to limit pressure build up during injection, natural \CO2 reservoirs may be closed systems. Understanding the dynamics of natural \CO2 accumulations is important, since they are our only analogs for long-term fate of geological \CO2 storage. 

{\color{black}{In the analyses presented below it should be kept in mind that the models presented here are based on numerous assumptions. Most importantly, our results are based on simulations in two-dimensional homogeneous isotropic systems with rectangular geometry, and they neglect hydrodynamic dispersion.}}

\subsection{Closed system dissolution in the Bravo Dome natural \CO2 field}
The Bravo Dome \CO2 field in New Mexico is commonly used as an analog for geological \CO2 storage, but recent work has shown that it comprises a number of isolated pressure compartments \citep{Akhbari2017}. It is unclear when these compartments became isolated and started acting as closed systems. In the calculation below we assume that they have been isolated for the majority of the lifetime of the reservoir. Here we focus on the NE-section of the reservoir, where significant \CO2 dissolution has occurred \citep{Gilfillan2009,Sathaye2014}. This section is separated from the main reservoir by a major fault and parts of it are underlain by a deep aquifer. In this section, the dissolution capacity is $\param\approx2$, the average depth of the reservoir is $H_w = 130$ m, the vertical permeability $\perm = 2.5\times10^{-15}$ m$^2$, the porosity $\varphi = 0.14$, the tortuosity of sandstone is $\tau \approx 4$, gravitational acceleration $g = 9.8$ m/s$^2$, initial density difference $\Delta\rho_0^* = 10.5$ kg/m$^3$, water viscosity at 35$^\circ$C is $\mu = 8.9\times10^{-4}$ Pa$\cdot$s, and the diffusivity of aqueous \CO2 is $D_m = 2\times10^{-9}$ m$^2$/s. 

Due to the relatively large tortuosity the effective diffusivity, $D=D_m/\tau$, in the sandstones is only $5\times10^{-10}$ m$^2$/s and may be even less in the lower porosity siltstones \citep{Hurlimann1994,Gist1990,Zecca2016}. The resulting initial Rayleigh number in this field is $Ra_0 \approx 540$ and the characteristic time scales are $\Tad \approx 4$ yrs, $\Ta\approx 2000$ yrs, and $\Td\approx 1$ Ma. {\color{black}{Although the Rayleigh number is large enough that Bravo Dome likely experienced convective \CO2 dissolution, it was not vigorous enough for a well-developed quasi-steady convection regime, so that the models developed in \S~\ref{sec:models} do not apply to Bravo Dome.}}


Instead, the diffusive models developed in \S~\ref{sec:Diffusion_Solution} provide an upper bound on the dissolution timescales at Bravo Dome. The time required for dissolved \CO2 to diffuse to the bottom of the reservoir is $t_d \approx 0.1\Td$, which corresponds to approximately 100,000 years in the NE-section of Bravo Dome. The time required to saturate the underlying aquifer is $t_d\approx\Td$ (figure~\ref{fig:DiffusionSoln}\emph{d}) and hence comparable to the estimated lifetime of the reservoir \citep{Sathaye2014}. This calculation assumes that all brine directly underlies the gas-water interface. In the NE-section of Bravo dome this is not strictly true, since the gas is localised within two domes  and significant lateral transport has to occur to saturate the entire brine within the reservoir. 

In the context of the simplified model explored here, however, it is possible that the underlying brine has been saturated. In this case, the fraction of \CO2 that has been dissolved at global equilibrium is given by
\begin{eqnarray}
    \frac{P_{g,0}^*-P_{g,e}^*}{P_{g,0}^*} = 1-C_{s,e} = \frac{\param}{1+\param} \approx 0.66.\label{eq:dissolved}
\end{eqnarray}
The maximum amount of dissolution in the NE-segment of Bravo Dome is limited to two-thirds of the amount that would have occurred in an equivalent open system. This is due to the significant drop in the gas pressure, which lowers the aqueous solubility of \CO2. The theoretical estimate (\ref{eq:dissolved}) is comparable to the estimate of 0.5, based on noble gases and reservoir characterisation \citep{Sathaye2014}. 

\begin{figure}[t]
{
  \centering
  \includegraphics[width=0.8\textwidth]{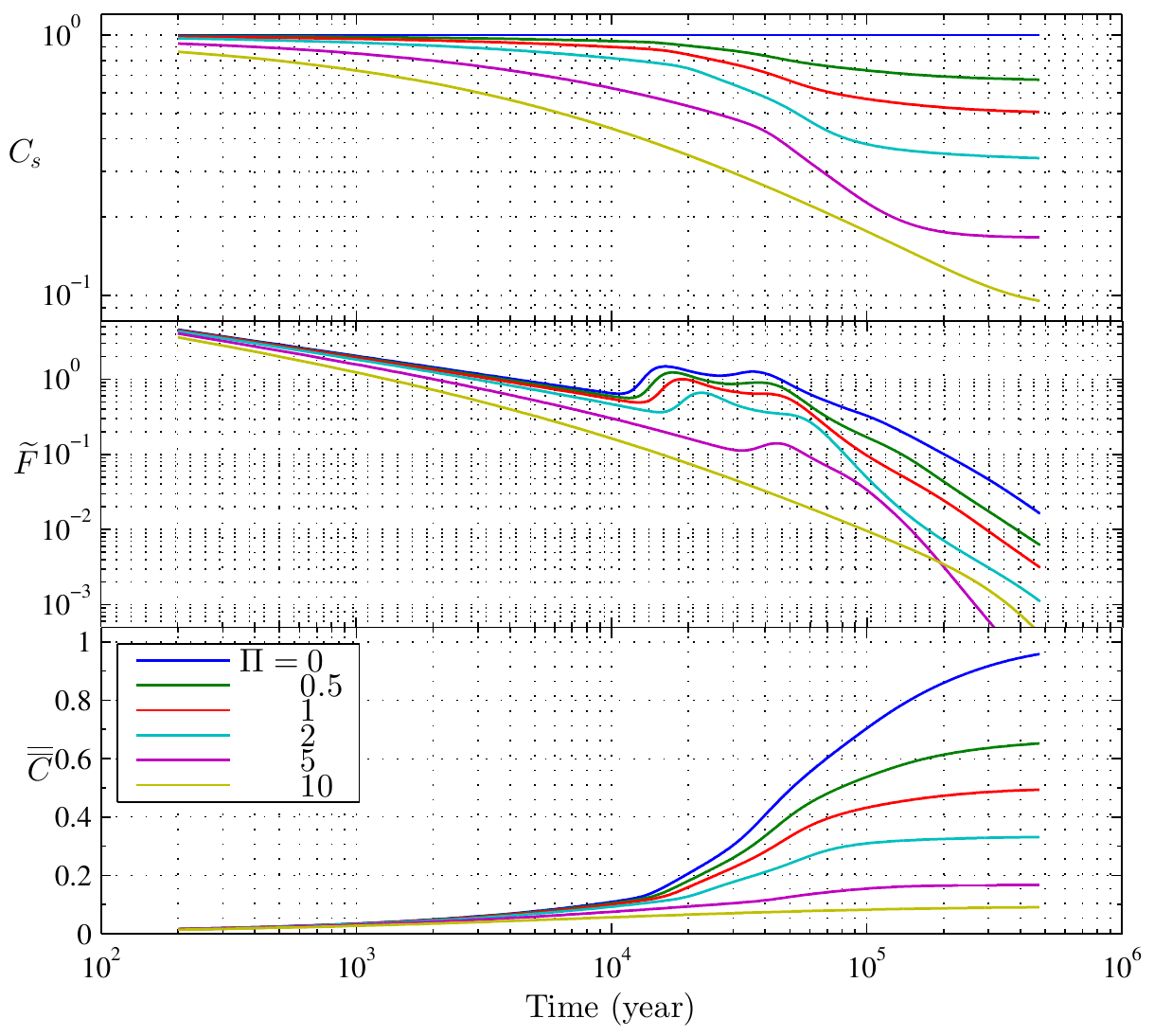}
}
\caption{{\color{black}{Evolution of the interface concentration $C_s$, the normalised dissolution flux $\widetilde{F} = F/(\alpha Ra_0)$ and the total dissolution $\overline{\overline{C}}$ in dimensional time for $Ra_0=540$, based on the Bravo Dome field parameters. The domain aspect ratio for the DNS reported here is $L = 10^5/Ra_0 \approx 185$.}}} \label{fig:DNS_Ra540}
\end{figure}



{\color{black}{To more comprehensively understand the convective \CO2 dissolution process in Bravo Dome, DNS were also performed at $Ra_0 = 540$ for various $\param$.  It is seen from figure~\ref{fig:DNS_Ra540} the \CO2 dissolution is significantly affected by $\param$ at moderate $Ra_0$: the onset time is delayed as $\param$ increases and at $\param \gtrsim 10$ no convection occurs, i.e. the transport is by diffusion. Based on these DNS data, for $\param = 0$ convection sets in after $130\,\Tad \approx 480$ yrs and the dissolution flux starts to grow after $3000\,\Tad \approx$ 11,000 yrs; while for $\param = 2$, the onset of convection occurs around $140\,\Tad \approx 520$ yrs and the dissolution flux starts to grow after $4300\,\Tad \approx$ 16,000 yrs. Moreover, figure~\ref{fig:DNS_Ra540} also reveals that no apparent quasi-steady convective regime exists at $Ra_0 = 540$, e.g. for $\param \le 5$ the convection begins to shut down right after the flux-growth \& plume-merging regime.  For $\param = 0$, the dissolution flux is halved after 70,000 years and is one-tenth of its initial value after 200,000 years; and the reservoir becomes 95\% saturated after 420,000 years. When system is closed, however, the dissolution flux declines significantly due to the negative feedback of pressure drop in the gas field: compared with $\param = 0$, the dissolution flux for $\param = 2$ is halved after 29,000 years with gas pressure reduced to 65\% and becomes one-tenth after 82,000 years with gas pressure reduced to 40\%; and the reservoir becomes 95\% saturated after 130,000 years. The closed system therefore saturates earlier than the open system, but the total amount dissolved is less due to the drop in gas pressure.}} When  comparing these estimates the simplifications in the model and the large uncertainties in the interpretation of the field data should be kept in mind.


\subsection{Timescales of high-Rayleigh-number convection in closed system}
To illustrate the effect of a closed system on a vigorously convecting system, we apply the high-$Ra$ convection models developed in \S~\ref{sec:models} to a hypothetical closed, high-permeability reservoir used in previous work \citep{Neufeld2010,Hewitt2013shutdown}. The parameters $H_w = 20$ m, $\perm = 2.5\times10^{-12}$ m$^2$, $\Delta\rho_0^* = 10.5$ kg/m$^3$, $\varphi = 0.375$, $D_m = 2\times10^{-9}$ m$^2$/s, and $\mu = 5.9\times10^{-4}$ Pa$\cdot$s are loosely based on the Sleipner site in North sea \citep{Bickle2007, Pau2010}. This reservoir comprises unconsolidated sand, so that $\tau \approx \sqrt{2}$ and the effective diffusivity is $D=D_m/\tau \approx 1.4\times10^{-9}$. The resulting initial Rayleigh number is $Ra_0 \approx 1.6\times10^4$, and the characteristic time scales are $\Tad \approx 0.3$ hr, $\Ta\approx 0.5$ yr, and $\Td\approx 9000$ yrs. As discussed in \S\ref{sec:Onset} and \S\ref{sec:DNSresults}, the dynamics in the diffusion-dominant regime are generally not affected by $\param$ at large $Ra_0$. Therefore, in such closed aquifers, convection sets in after $130\,\Tad \approx 2$ days and the dissolution flux starts to grow after $3000\,\Tad \approx 36$ days. 

To exhibit the long-term effect of the parameter $\param$ on \CO2 dissolution, we estimate the evolution of the \emph{normalised} gas pressure $P_g^*/P_{g,0}^*$ (i.e. $C_s$), dissolution flux $F/(\alpha Ra_0)$ and total dissolved \CO2 (i.e. $\overline{\overline{C}}$) in time for this high-permeability reservoir using the box models developed in \S~\ref{sec:models}. As shown in figure~\ref{fig:Sleipner}, for $\param = 0$ the convection starts to shut down after 9 years; the dissolution flux is halved after 19 years and is one-tenth of its initial value after 60 years; and the reservoir becomes 95\% saturated after 330 years. For $\param = 2$, however, the pressure in the gas field declines significantly as \CO2 dissolves into the water: the convection shuts down after $11$ years, when the gas pressure is reduced to 57\% of its initial value. Due to this negative feedback, the dissolution flux is halved after 7 years and is one-tenth of its initial value after 20 years, and the reservoir becomes 95\% saturated after 110 years.  


\floatsetup[figure]{style=plain,subcapbesideposition=top}  
\begin{figure}[t]
{
  \centering
  \quad\quad\;\;\includegraphics[width=0.67\textwidth]{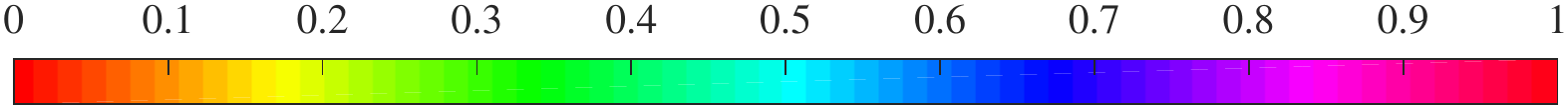} \vspace{-0.1in}\\
  \sidesubfloat[]{\includegraphics[width=0.7\textwidth]{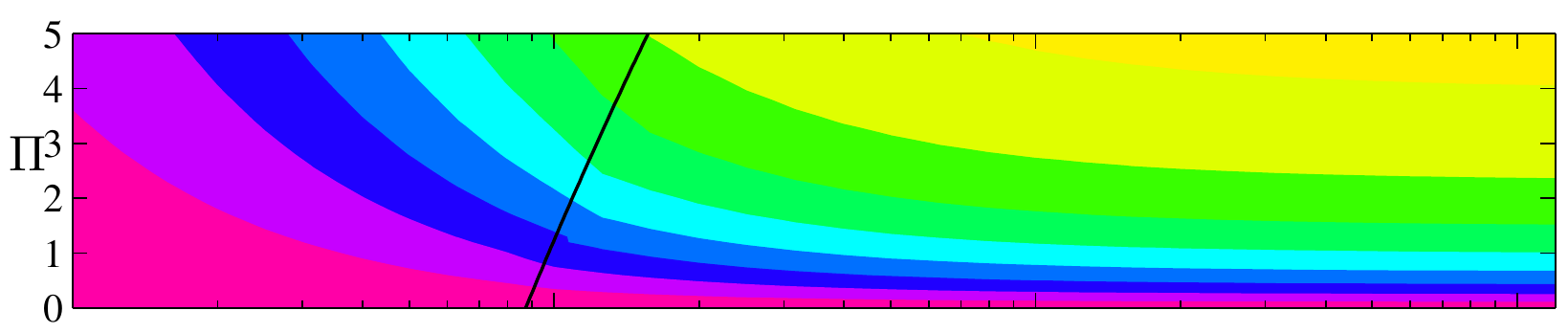}}\\
  \sidesubfloat[]{\includegraphics[width=0.7\textwidth]{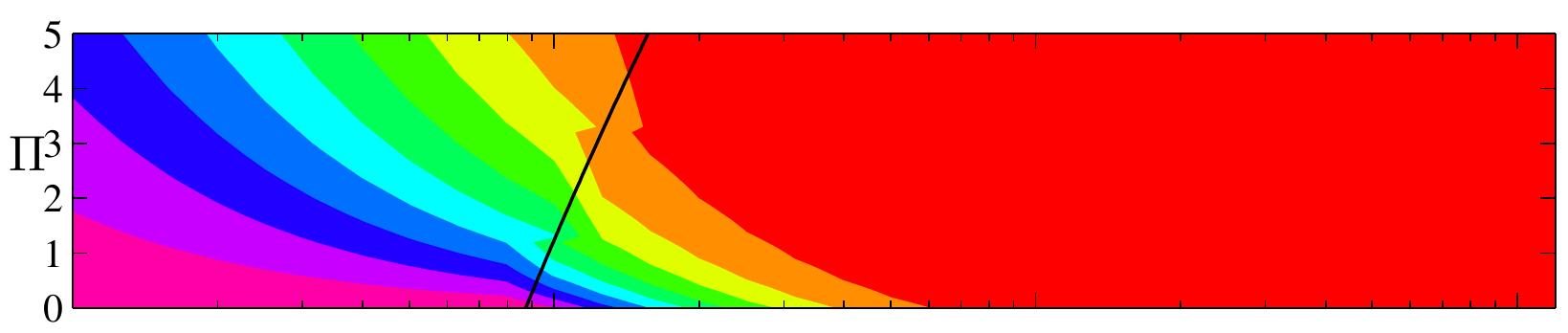}}\\
  \sidesubfloat[]{\includegraphics[width=0.7\textwidth]{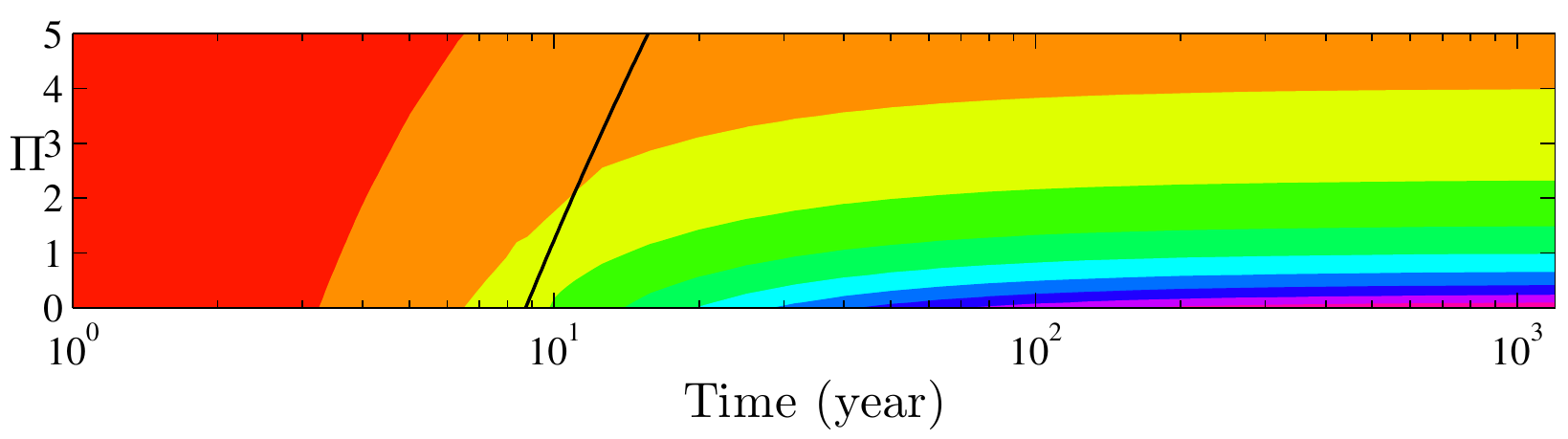}}\par
}
\caption{{\color{black}{Contours of ($a$) the normalised gas pressure, $\widetilde{P} = P_{g}^*/P_{g,0}^* = C_s$, ($b$) the normalised dissolution flux, $\widetilde{F} = F/(\alpha Ra_0)$, and ($c$) the total dissolution $\overline{\overline{C}}$}}, based on the box models developed in \S~\ref{sec:models}, for the high-permeability reservoir.  The solid line marks the dimensional shut-down time $t_s^* = t_s\cdot \Ta$ based on (\ref{ts_model}). The zigzag contours around $t^* = t^*_s$ are due to the discontinuities of the box models for the quasi-steady convective and shut-down regimes at $t_s$.}  \label{fig:Sleipner}
\end{figure}

\section{Conclusions}\label{sec:Conclusion}
We have examined the dynamics of convective \CO2 dissolution in a closed porous media system, where the dissolution is accompanied by a drop in gas pressure. This introduces a negative feedback that slows both diffusive and convective mass transport and reduces the overall amount of \CO2 that can be dissolved. The strength of this negative feedback is controlled by the dimensionless dissolution capacity, $\param$, which corresponds to the fraction of the initial gas that can be dissolved into the water at equilibrium. The dynamics in a closed system, $\param > 0$, differ fundamentally from those in an open system, since the interface concentration, which drives mass transport, declines with time. In closed systems diffusive mass transport is no longer self-similar and convective mass transport is never quasi-steady with a constant flux. However, we use DNS to show that the flux, $F$ is quadratic in the interface concentration $C_s$ at high Rayleigh numbers. This allows the construction of box models that successfully capture the mean behavior of the convecting system. Our results show that the pressure drop in closed systems can significantly limit convection long before the underlying brine begins to saturate. This may explain the persistence of natural \CO2 accumulations in isolated reservoir compartments over geological time periods. 

\begin{acknowledgments}
This work was supported as part of the Center for Frontiers in Subsurface Energy Security, an Energy Frontier Research Center funded by the U.S. Department of Energy, Office of Science, Basic Energy Sciences under Award \# DE-SC0001114. B.W. acknowledges a postdoctoral fellowship through the Institute of Computational and Engineering and Science at the University of Texas at Austin. 
\end{acknowledgments}

\bibliographystyle{jfm}

\bibliography{jfm-ClosedSystem}

\end{document}